\documentclass[iop,revtex4-1]{emulateapj}

\usepackage{epsfig}
\usepackage{epstopdf}
\usepackage{graphics}

\slugcomment{To appear in The Astrophysical Journal Supplement Series.}

\shorttitle{Methyl Cyanide Observations Toward Massive Protostars}
\shortauthors{Rosero et al.}

\begin{document}

\title{Methyl Cyanide Observations Toward Massive Protostars} 
\bigskip\bigskip

\author{V. Rosero$^{1}$, P. Hofner$^{1,}$\altaffilmark{\dag}, S. Kurtz$^{2}$,
J. Bieging$^{3}$ \& E. D. Araya$^{4}$ }

\affil{$^1$Physics Department, New Mexico Tech, 801 Leroy Pl., Socorro, NM 87801, USA}

\affil{$^2$Centro de Radioastronom{\'\i}a y Astrof{\'\i}sica, 
Universidad Nacional Aut\'onoma de M\'exico, Morelia 58090, M\'exico}

\affil{$^3$Department of Astronomy and Steward Observatory, University of Arizona, 933 North Cherry Avenue, Tucson, AZ 85721, USA}

\affil{$^4$Physics Department, Western Illinois University, 1 University Circle, Macomb, IL 61455, USA}

\altaffiltext{\dag}{Adjunct astronomer at the National Radio Astronomy Observatory.}

\begin{abstract}

{We report the results of a survey in the  CH$_3$CN J\,=\,12 $\rightarrow$ 11 transition toward 
a sample of massive proto-stellar candidates. The observations were carried out with the $10\,$m Submillimeter telescope on Mount Graham, AZ.
We detected this molecular line in 9 out of 21 observed sources. In six cases this is the first detection of this transition.
We also obtained full beam sampled cross-scans for five sources which show that the lower $K-$components can be extended
on the arcminute angular scale. The higher $K-$components however are always found to be compact with respect to our
$36^{\prime\prime}$ beam. A Boltzmann population diagram analysis of the central spectra indicates CH$_3$CN column densities of
about $10^{14}\,$cm$^{-2}$, and rotational temperatures above $ 50\,$K, which confirms these sources as hot molecular cores.
Independent fits to line velocity and width for the individual $K-$components resulted in the detection of an increasing blue shift with increasing
line excitation for four sources. Comparison with mid-infrared images from the \emph{SPITZER GLIMPSE}/IRAC archive for six sources  show that the
CH$_3$CN emission is generally coincident with a bright mid-IR source. Our data clearly show that the CH$_3$CN J\,=\,12 $\rightarrow$ 11
transition is a good probe of the hot molecular gas near massive protostars, and provide the basis for future interferometric studies.}\end{abstract}


\section{Introduction}

Despite the central role of massive stars in almost all areas of astronomy, the physical processes involved in the formation of stars with masses $> 8\,$M$_\odot$
are at present 
poorly understood (e.g., \citet{2007ARA&A..45..481Z}). Theoretical and observational studies favor the idea that massive stars, similar to their lower mass counterparts,
form from a collapsing cloud core. However, whether the subsequent formation of a protostar and further mass accretion occurs from an isolated massive
molecular core (e.g., \citet{2003ApJ...585..850M}, \citet{2007ApJ...666..976K}), or under the influence of a cluster environment (e.g., \citet{2004MNRAS.349..735B}) remains a persistent question.

\begin{deluxetable*}{lrrcccccr}[h!]		  		  
\tabletypesize{\scriptsize}					  
\tablecaption{Observed Sources}  
\tablewidth{0pt}				  						  
\tablehead{					  
\colhead{Source} &				  
\colhead{R.A.} &				  
\colhead{Dec.} &				  
\colhead{V$_{LSR}$} &
\colhead{d} &
\colhead{L} &
\colhead{rms} &
\colhead{CH$_{3}$CN}&
\colhead{Ref.\tablenotemark{a} }
\\	[3pt] 				  
\colhead{} &					  
\colhead{(J2000)} &					  
\colhead{(J2000)} &	
\colhead{(km\,s$^{-1}$)} &
\colhead{(kpc)} &	
\colhead{($10^{4}$ L$_{\odot}$)} &
\colhead{(mK)} &	
\colhead{Detection} & }						  
						  
\startdata                                       
IRAS 18151$-1208$ & 18:17:58.2 & $-12$:07:26        & 33.0 & 3.0   & 2 & 20 & n & (1)  \\	[3pt] 
G16.59$-0.05$ & 18:21:09.0 & $-14$:31:49                  & 60.0 & 4.5   & 2\tablenotemark{b}  & 6 & y& (1)(2)\\	[3pt] 
IRDC 18223$-3$ & 18:25:08.3 & $-12$:45:27              & 45.0 &  3.7  &0.018  & 20 & n & (1)(3) \\	[3pt] 
IRAS 18264$-1152$ & 18:29:14.4 & $-11$:50:24        & 45.0 & 3.5   & 1\tablenotemark{b} & 6 & y & (1) \\	[3pt] 
G23.01$-0.41$ & 18:34:40.4 & $-09$:00:40                  & 80.0 & 4.6   & 20& 6 & y& (4)(5)(6)\\	[3pt] 
G23.71$-0.20$ & 18:35:12.4 &$-08$:17:39                   & 79.2 & 11.0   & 24\tablenotemark{c} &20&n & (7)(8)\\	[3pt] 
G25.83$-0.18$ & 18:39:03.6 & $-06$:24:11                  & 91.0 &  5.6 & \nodata & 8 & y & (9)\\	[3pt] 
G28.87+0.07 & 18:43:46.0 & $-03$:35:30                     & 105.0& 7.4 & 19 & 7 & y & (2)\\	[3pt] 
G34.26+0.15 & 18:53:18.5 & 01:14:58                            & 58.8 & 3.7 & 47 &20& y & (10)(11)\\	[3pt] 
IRAS 18566+0408 & 18:59:10.0 & 04:12:16                  & 85.5 & 6.7  & 6.3 & 7 & y & (1)\\	[3pt] 
IRAS 20126+4104 & 20:14:26.0 & 41:13:32                 & $-3.5$& 1.7 & 4 & 6 & y & (12)(13)\\	[3pt] 
IRAS 21307+5049 & 21:32:30.6 & 51:02:17                  &$-47.0$& 3.6 & 0.4 &14& n & (13) (14)\\	[3pt] 
IRAS 21519+5613 & 21:53:38.8 & 56:27:49                  & $-63.0$& 7.3 & 1.9 &20& n & (15)\\	[3pt] 
IRAS 22172+5549 & 22:19:08.6 & 56:05:02                  & $-44.0$& 2.4 & 0.2 &20& n & (14) \\	[3pt] 
IRAS 22506+5944 & 22:52:38.7 & 60:00:56                  & $-51.5$& 5.7  &2.2 &20& n & (15)\\	[3pt] 
IRAS 23033+5951 & 23:05:24.8 & 60:08:14                  & $-53.0$& 3.5   & 1 &20& n & (1)\\	[3pt] 
NGC 7538$\:$S & 23:13:45.0 & 61:26:49                       & $-56.0$& 2.8    & 1.6  & 6 & y & (16)(17) \\	[3pt] 
NGC 7538$\:$IRS$\:$9 & 23:14:01.7 & 61:27:20         & $-57.0$& 2.8    & 2 &20& n & (16)(17)(18) \\	[3pt] 
IRAS 23139+5939 & 23:16:10.3 & 59:55:28                  & $-44.5$&  4.8  & 2.5 &20& n & (1)\\	[3pt] 
IRAS 23151+5912 & 23:17:21.0 & 59:28:49                   & $-54.0$& 5.7   &10 & 20 & n &(1)\\	[3pt] 
IRAS 23385+6053 & 23:40:54.5 & 61:10:28                  & $-50.0$&  4.9   &1.6& 20 & n & (14)\\	[3pt] 
\enddata
\tablenotetext{a}{References for  distance and luminosity values.}	
\tablenotetext{b}{Source has a distance ambiguity. Reported luminosity corresponds to the near distance.}	
\tablenotetext{c}{Source has a distance ambiguity. Reported luminosity corresponds to the far distance.}
\tablecomments{\footnotesize{Units of right ascension are hours, minutes, and seconds, and units of declination are degrees, arcminutes, and arcseconds. 
\\
{\bf References.} (1) \citet{2002ApJ...566..931S}; (2) \citet{1997A&A...325..282C}; (3) \citet{2007ApJ...656L..85B}; (4) \citet{2009ApJ...693..424B}; (5) \citet{2008ApJS..178..330A}; (6) \citet{2005A&A...432..737P}; 
(7) \citet{2004ApJS..154..553S};
(8) \citet{2006ApJ...643L..33A}; (9) \citet{2006MNRAS.367..553P}; (10) \citet{1990A&AS...83..119C}; (11) \citet{1999ApJS..125..143W}; (12) \citet{2000ApJ...535..833S}; (13) \citet{2008ApJ...681..415T}; (14) \citet{2002ApJ...570..758M};  
(15) \citet{2004ApJ...604..258S}; (16)  \citet{1979MNRAS.188..463W}; (17)   \citet{2004ApJ...600..269S}; (18) \citet{1979ApJ...230..133T}}} \label{t1}

\end{deluxetable*} 

One of the earliest observational manifestations of massive proto-stars are so-called hot molecular cores (hereafter HMCs). Named after the prototype object in the
Orion KL region, they have been surveyed throughout the Galaxy in an effort to identify candidates for massive proto-stars (e.g., \citet{2002ApJ...566..931S}).
Common search criteria were high FIR luminosity, high molecular column densities and temperature, and the absence of strong radio continuum emission --- the latter to ensure
an earlier evolutionary phase than ultra-compact (UC) or hyper-compact (HC) HII regions.

A particularly useful tracer of HMCs is the methyl cyanide (CH$_3$CN) molecule.  Due to the centrifugal deformation of this symmetric top molecule, its rotational spectrum
consists of a series of closely spaced $K-$components tracing rapidly increasing excitation energies.  The $K-$ladders are connected only through collisions so that
excitation temperatures can in principle be measured from the ratios of $K-$components, thus avoiding the usual calibration uncertainties that occur
when comparing rotational
transitions observed in different frequency bands. An extensive discussion of the microwave spectroscopy of methyl cyanide is given in \citet{1980JPCRD...9..659B}.
Because of these spectroscopic properties, the CH$_3$CN molecule is frequently used to determine
temperatures in the dense molecular cores where massive stars form, using both statistical equilibrium calculations (e.g., \citet{1984ApJ...286..232L}), or the simpler
rotation diagram technique (e.g., \citet{1999ApJ...517..209G}).                                                                                                                                                                                                                                                                                                                                                                                                                                                                                                                                                                                                                                                                                                                                                                                                                                                                                                                                                                                                                                   

Another factor that favors the use of CH$_3$CN as a tracer of HMCs is its enhanced abundance in warm (T$\, = 100 - 300\,$K), dense
($n_{H_2} = 10^6 - 10^8\,$cm$^{-3}$) environments (e.g., \citet{1987ApJ...315..621B}). This is generally thought to be caused by grain surface chemistry, either by primary reactions
on the grain surface with subsequent release into the gas phase when the grain mantle evaporates, or, alternatively, by secondary reactions in the gas phase 
(e.g., \citet{1992ApJ...399L..71C}, \citet{2008A&A...488..959B}). Recently, \citet{2009A&A...507L..25C} reported detection of CH$_3$CN in the outflow lobes of the low-mass protostar 
L1157-B1, and attributed the enhanced CH$_3$CN abundance to shock chemistry. From these studies it is well-established that CH$_3$CN traces an 
energetic environment similar to that expected in the immediate vicinity of massive proto-stars.

To investigate the HMC phase of massive star formation, 
several single dish CH$_3$CN surveys have been made (e.g., \citet{1993A&A...276..489O}, \citet{2005ApJS..157..279A}, \citet{2001ApJ...558..194P}), 
and a small number of sources have also been studied with mm-interferometers (e.g., \citet{1994ApJ...435L.137C}, \citet{1996ApJ...460..359H}, \citet{2008ApJ...673..363F}).
Several of the interferometric studies resulted in images of CH$_3$CN structures that are elongated perpendicular to the direction of molecular outflows,
with velocity gradients
along the elongated structures; this is usually explained as rotational motion of a circumstellar disk or torus (e.g., \citet{2011A&A...525A.151B}). The CH$_3$CN
molecule is thus a good choice to trace accretion disks around massive proto-stars, whose existence (if confirmed), and properties will be important input for
current models of massive star formation.

From the above discussion it is clear that observations of CH$_3$CN with high sensitivity and angular resolution are well-suited to study accretion disks
around massive stars. Such observations have recently become more accessible using instruments such as the Submillimeter Array  (SMA) and ALMA. To facilitate such observations in the 
CH$_3$CN J\,=\,12 $\rightarrow$ 11 line, the present study adds to the existing database of single dish studies of methyl cyanide.

In Section \ref{observations_sec} we describe the observations and data reduction, and we present the 
observational results in Section \ref{results_sec}.  We conclude in Section \ref{disc_sec}  with a discussion of these observational results.

\section{Observations and Data Reduction}\label{observations_sec}

We observed 21 sources in 2008 from May 26 to 31 in the 1.3 mm CH$_3$CN J\,=\,12 $\rightarrow$ 11 
transition  with the 10 m Heinrich Hertz Submillimeter Telescope (SMT)\footnote{Operated by the Arizona Radio Observatory
(ARO) at the University of Arizona}  on Mt. Graham,  AZ.  The telescope beam width at $1.3\,$mm is
approximately 36$\arcsec$, and the pointing accuracy during our observing run
was better than 7$\arcsec$. The pointing positions and LSR velocities 
for the observed sources are given in Table~1.
Most of our target sources are prominent HMC candidates with large IRAS luminosities. 
They all show the typical observational indicators of massive star formation in the HMC stage, namely
massive molecular cores, H$_2$O and CH$_3$OH maser emission, warm molecular gas, weak (or absent) radio continuum
emission, and the presence of jets and molecular flows. 
An  exception is IRDC 18223-3, 
which is a massive infrared dark cloud with an embedded protostar, which may be in an 
earlier evolutionary state \citep{2007ApJ...656L..85B}. 

The observations were conducted in double sideband mode using the 1.3 mm J--T ALMA sideband separating receiver\footnote{
http://aro.as.arizona.edu/smt\_docs/receivers/1.3mm\_ALMA/Manual.pdf}, which simultaneously recorded two linear orthogonal polarizations.
The CH$_3$CN J\,=\,12 $\rightarrow$ 11 $K=5$ transition ($\nu_0$ = $220.641089\,$ GHz, Boucher et al. 1980) 
was tuned to the center of the lower sideband. We used all six available backends;
four low spectral resolution and two high spectral resolution. The low spectral resolution
backends were the acoustic-optic spectrometers (AOSs) AOS-A and AOS-B with bandwidths and spectral 
resolutions of $975\,$ MHz ($1325\,$km$\,$s$^{-1}$) and $953\,$kHz 
($1.3\,$km$\,$s$^{-1}$), and 
 two filterbanks which have bandwidths and spectral resolutions 
of 1.024$\,$GHz ($1392\,$km$\,$s$^{-1}$) and $1\,$MHz ($1.4\,$km$\,$s$^{-1}$),
respectively.  Most of the data used for the analysis in this paper were taken 
with the AOS-C and the 
Chirp Transform Spectrometers (CTS-A),  which have bandwidths and channel widths
 of $250.5\,$MHz ($340\,$km$\,$s$^{-1}$) and $122\,$kHz 
($0.17\,$km$\,$s$^{-1}$) and $215\,$MHz ($292\,$km$\,$s$^{-1}$) and $29\,$kHz
 ($0.04\,$km$\,$s$^{-1}$), respectively. 
 
The AOS-C bandwidth does not cover all $K-$components of the CH$_3$CN J\,=\,12 $\rightarrow$ 11 transition, hence we 
centered this backend on the $K=5$ component. This setup allowed us to 
observe the $K=0 -7$ components simultaneously. The AOS-A and AOS-B and the filterbanks 
offer  bandwidths between $~950$ and $1000\,$MHz  so in principle
the higher $K-$components could be detected. In practice, however, the lower line strengths
of the higher components prevented us from detecting them in all but the strongest source,
G34.26+0.15. Furthermore, the $^{13}$CO J\,=\,2 $\rightarrow$ 1 transition 
blends with the $K=9$ component of the CH$_3$CN J\,=\,12 $\rightarrow$ 11 
transition, thus limiting the usefulness of this $K-$component. 

The $^{12}$CO J\,=\,2 $\rightarrow$ 1 transition located in the upper sideband (USB)
 contaminated the lower sideband data approximately 50 MHz away from the 
CH$_3$CN J\,=\,12 $\rightarrow$ 11 $K=0$ line.  With the exception of
G28.87+0.07, this contamination  had little effect on the quality of our data. 
The USB rejection was approximately $17\,$dB  over 
the course of our observations.  Given the relatively weak appearance of the
$^{12}$CO J\,=\,2 $\rightarrow$ 1 line (normally a very strong line) in our spectra,
we do not believe that there is significant contamination from other spectral lines located in the 
USB.  
 
Our observations were conducted using double beam switching with a switch rate of 
$2\,$Hz and a beamthrow of 2\arcmin\
with a total on/off cycle of approximately 6 minutes per scan.
System temperatures ranged from 350\,K to just under 
200\,K with an average temperature of 212\,K. Focus corrections were obtained from
observations of Jupiter.
Whenever possible we also derived pointing corrections from cross scans of Jupiter.
When Jupiter was not available or was located at a large angular distance from the 
target source, pointing corrections were made by observing asymptotic giant branch stars 
in the CO J\,=\,2 $\rightarrow$ 1 transition falling in the USB. In 
the case of IRAS 20126+4104 we could line point using the
$^{13}$CO J\,=\,1 $\rightarrow$ 0 emission from the source. 

At the beginning of each night of observations, we observed the 
strong source G34.26+0.15 for at least one scan to check for day-to-day consistency
of our observations and to obtain a template source to identify contaminating spectral lines.
Subsequently, all sources were observed for 12--18 minutes to determine the 
intensity of the $K-$components of the CH$_3$CN J\,=\,12 $\rightarrow$ 11 transition.
Promising sources were then re-observed for at least  2 hr.
Additionally, we obtained full beam spaced cross scans for five sources,
typically with  $1.5\,$hr spent at each offset position.

The data were reduced in CLASS, which is part of the GILDAS{\footnote{http://www.iram.fr/IRAMFR/GILDAS}} software package.
All spectra were first inspected  to check for  bad channels or 
any obvious artifacts; bad scans were discarded.
Subsequently, we subtracted baselines using low order polynomials, and initially averaged the spectra for each
spectrometer and each day separately. After further inspection, all data taken on different days
were averaged to form a final data set for each source. 
After Hanning smoothing and resampling to the same spectral resolution, we 
averaged the AOS-C and CTS-A spectrometer data. We will refer to the latter data set as `high resolution' spectra.
The data were calibrated using the chopper-wheel method and the antenna temperature was
converted to main-beam brightness temperature by dividing the antenna temperature
by the main-beam efficiency of the telescope ($\eta_{b}=0.74$){\footnote{http://aro.as.arizona.edu/smt\_docs/smt\_efficiency.pdf}}.

Using the daily spectra of the strong source G34.26+0.15, we checked our data for
amplitude stability, which maximum deviation from the average was found to be 
smaller than  $16\,\%$. Measured line widths had
maximum deviations of $13\,\%$, and the repeatability of measured frequencies, as well as the
linearity of the spectrometers, was better than $1\,\%$.

We have three common sources
with \citet{2001ApJ...558..194P}, who observed the same CH$_{3}$CN transition with the SMT, albeit with a different receiver.
These three sources are G34.26+0.15, IRAS 23139+5939 and IRAS 23385+6053, the latter 
two being non-detections by us as well. The results of our line fitting of 
G34.26+0.15 agree very well with the spectrum of the same source shown in \citet{2001ApJ...558..194P}. 

\section{Results}\label{results_sec}

\subsection{Line Contamination}\label{line_cont}

\begin{figure*}
\centering
\includegraphics[scale=0.70]{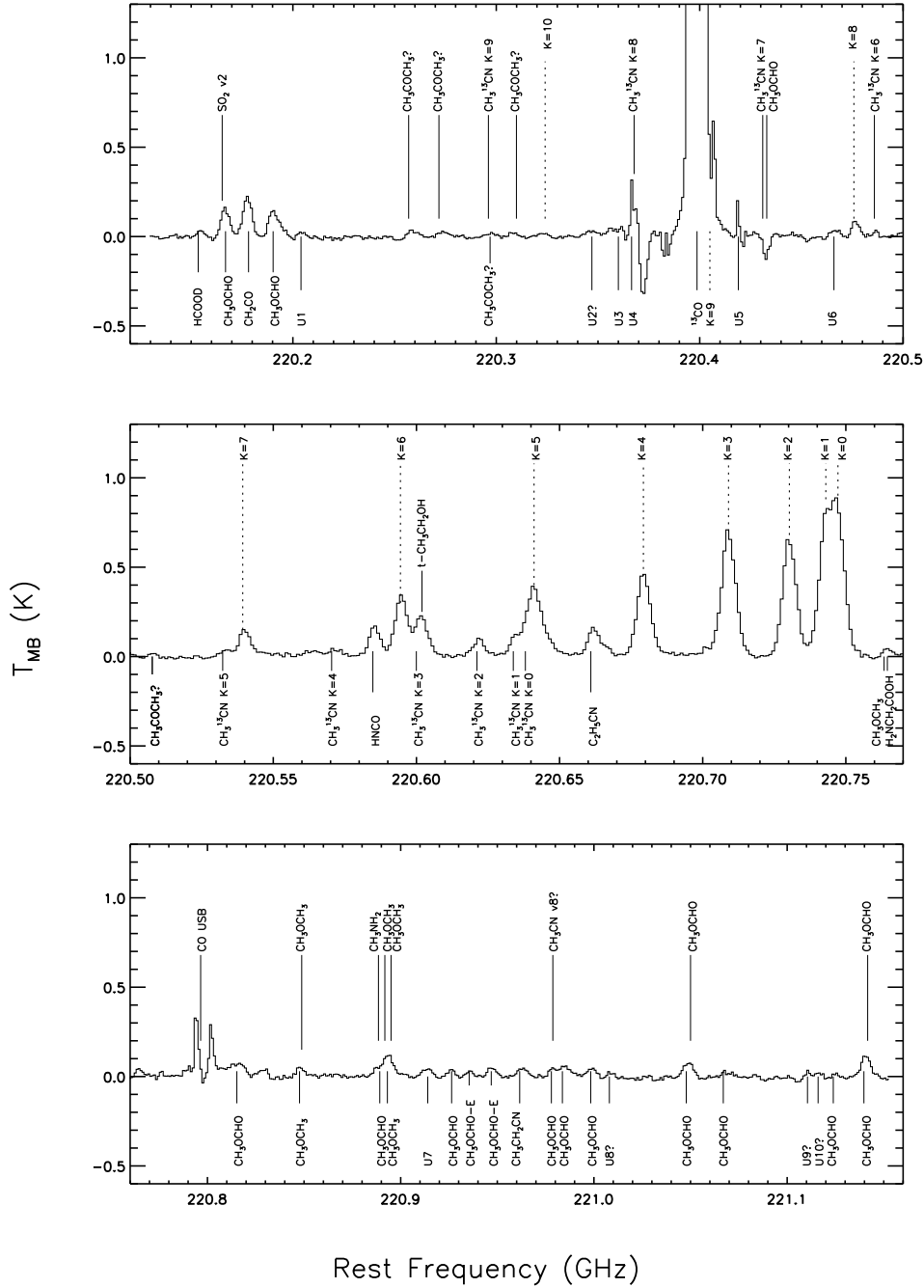}
\caption{The above spectrum shows the full bandwidth of the combined AOS-A and B spectrometers for the 
source G34.26+0.15. The spectral resolution is about $1\,$MHz. Line identifications are indicated by vertical lines.}\label{f1}
\end{figure*}

As mentioned above, we obtained spectra of the strong source G34.26+0.15 to study
possible line contamination of the CH$_3$CN J\,=\,12 $\rightarrow$ 11 transition.
In Figure \ref{f1} we show the full $1\,$GHz bandpass for this source. 
In addition to the $K-$components of the CH$_3$CN J\,=\,12 $\rightarrow$ 11 
transition many other molecular
lines were detected. We used the JPL Molecular Spectroscopy Catalog \citep{1998JQSRT..60..883P} in conjunction with the 
Cologne Database for molecular spectroscopy 
(CDMS, \citet{2001A&A...370L..49M}) to identify the detected lines.

In some cases there was a high level of ambiguity
in the identities of the spectral lines. In these cases, molecules that were unambiguously detected
elsewhere in the bandpass  were preferentially chosen.   Lines that we were unable to identify
are labeled as ``U'' for unknown lines and are numbered according to their order of appearance.
Over 60 lines were identified across
the full bandpass and we have detections in the CH$_3$CN J\,=\,12 $\rightarrow$ 11 transition
up to $K=10$. The same transitions in the isotopologue
CH$_{3}^{13}$CN were also detected, possibly out to $K=9$. In the case of this isotopologue
 the $K=0, 1$ and 3 components and the $K=7$ and higher components are blended with other
lines. Absorption features are observed around the $^{13}$CO J\,=\, 2 $\rightarrow$ 1 line; these features
are possible artifacts arising from $^{13}$CO emission in the off-beam, and the line identifications in this
area of the spectrum are tentative at best.

From inspection of Figure \ref{f1}, we conclude that the CH$_3$CN J\,=\,12 $\rightarrow$ 11
$K= 0$ -- 4 , 7, 8, and 10 suffer no significant contamination
but the $K= 5$, 6 components suffer some blending with lines from other molecular species. 
The $K=9$ component is rendered useless by overlap with the $^{13}$CO$(2-1)$ line.   
In the analysis that follows we use the $K= 5$, 6 components only when we are confident that we can
reasonably separate them from the other lines; the $K=9$ component we do not use at all.

\subsection{CH$_3$CN Line Detection and Analysis}\label{pop_diag_sec} 

\begin{figure*}
\centering
\includegraphics[scale=0.85]{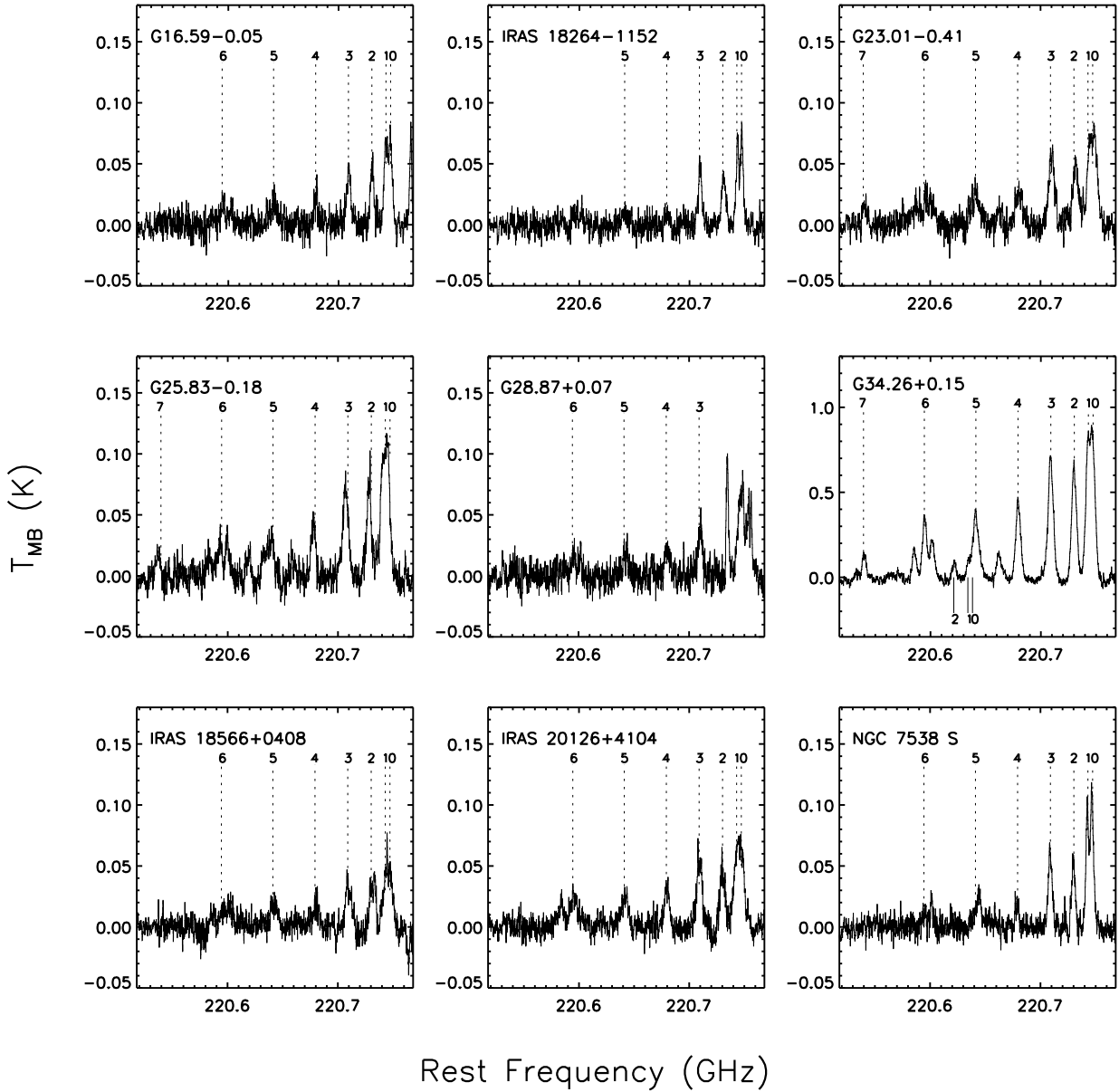}
\caption{Spectra at the central position for the sources with
CH$_3$CN J\,=\,12 $\rightarrow$ 11 detections. The dashed lines indicate $K-$components for the main 
isotopologue. 
In the spectrum for the source G34.26$+$0.15 we also indicate the detected $K-$components of  
CH$_3$$^{13}$CN with solid lines. USB emission is affecting the $K=0$, 1, 2 lines in G28.87$+$0.07.}\label{f2}
\end{figure*}

 We detected emission in the CH$_3$CN J\,=\,12 $\rightarrow$ 11 transition towards 9 of the 21 sources 
 of the sample (see Table \ref{t1}). In Figure \ref{f2} we show our high resolution spectra at the center position. $K-$components were 
 detected up to $K=7$ for several sources. The upper excitation energy of the $K=7$ component is about $420\,$K \citep{1980JPCRD...9..659B}, thus indicating the 
 presence of hot molecular gas.

We have performed simultaneous Gaussian fits for all $K-$components of the CH$_3$CN J\,=\,12 $\rightarrow$ 11 transition.
The FWHM of  all lines was kept constant at the value measured for the unblended $K=3$ component, and the relative
position of all lines was fixed at the theoretical values.
For the main isotopologue the $K=5$ line was fit with 3 Gaussians when blending with CH$_3$$^{13}$CN $K=0$ and $K=1$ was 
suspected. 
The $K=6$ line was fit taking into account that its high-frequency wing is blended with the $K=3$ component of CH$_3$$^{13}$CN.

The magnitude of the hyperfine splitting (hfs) of the CH$_{3}$CN molecule reported in \citet{1980JPCRD...9..659B} for the CH$_3$CN J\,=\,12 $\rightarrow$ 11
 transition is $\le 0.3\,$MHz. Due to the large FWHM ($\sim 8$ MHz), and the relatively low signal-to-noise
 detections of the higher $K-$components, we omitted the hfs in the line fitting.   
 In Table \ref{t2} we list the line parameters from the Gaussian fits.
 
\begin{figure*}
\centering
\begin{tabular}{cc}
\epsfig{file=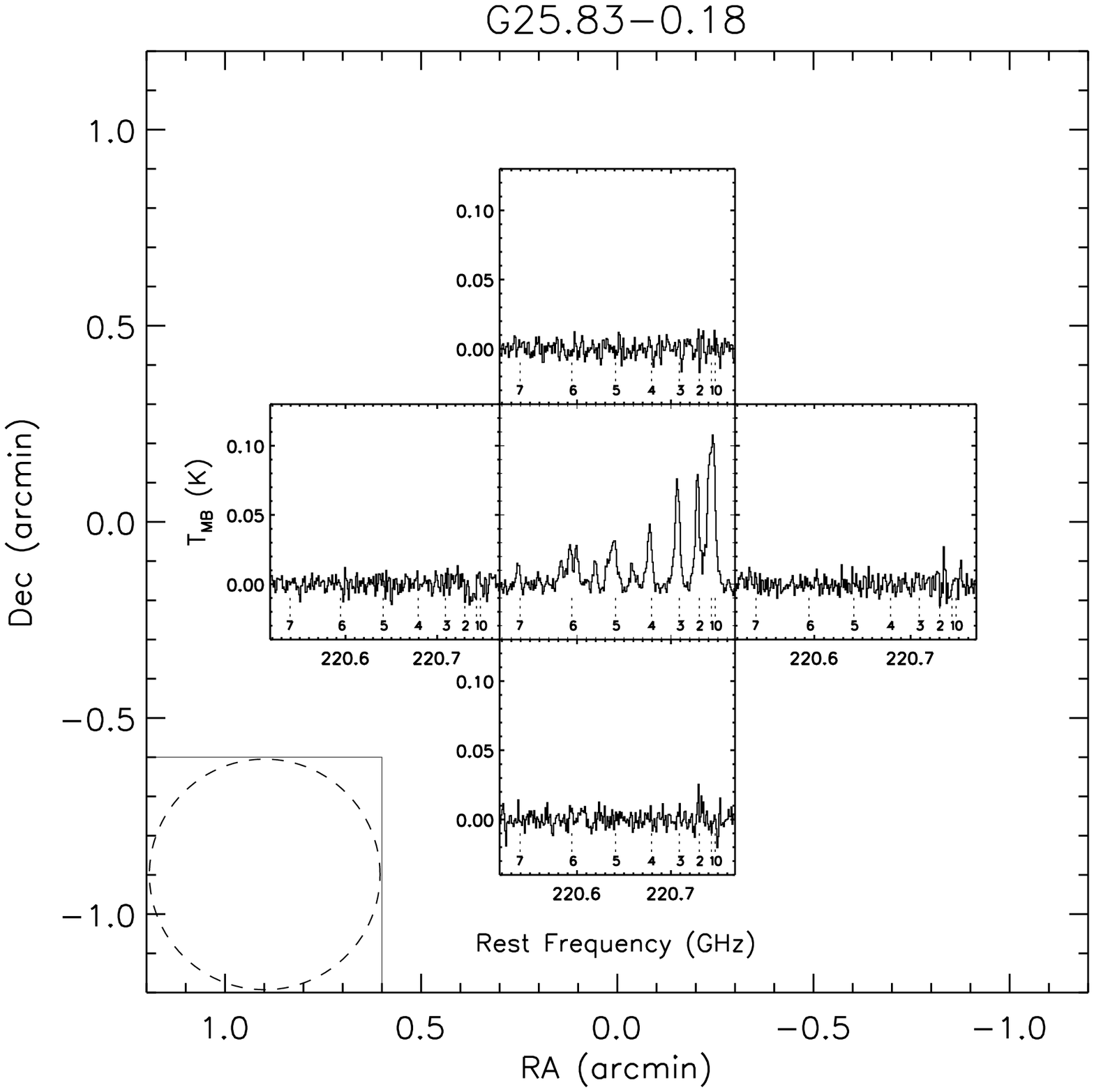,width=0.41\linewidth} & 
\epsfig{file=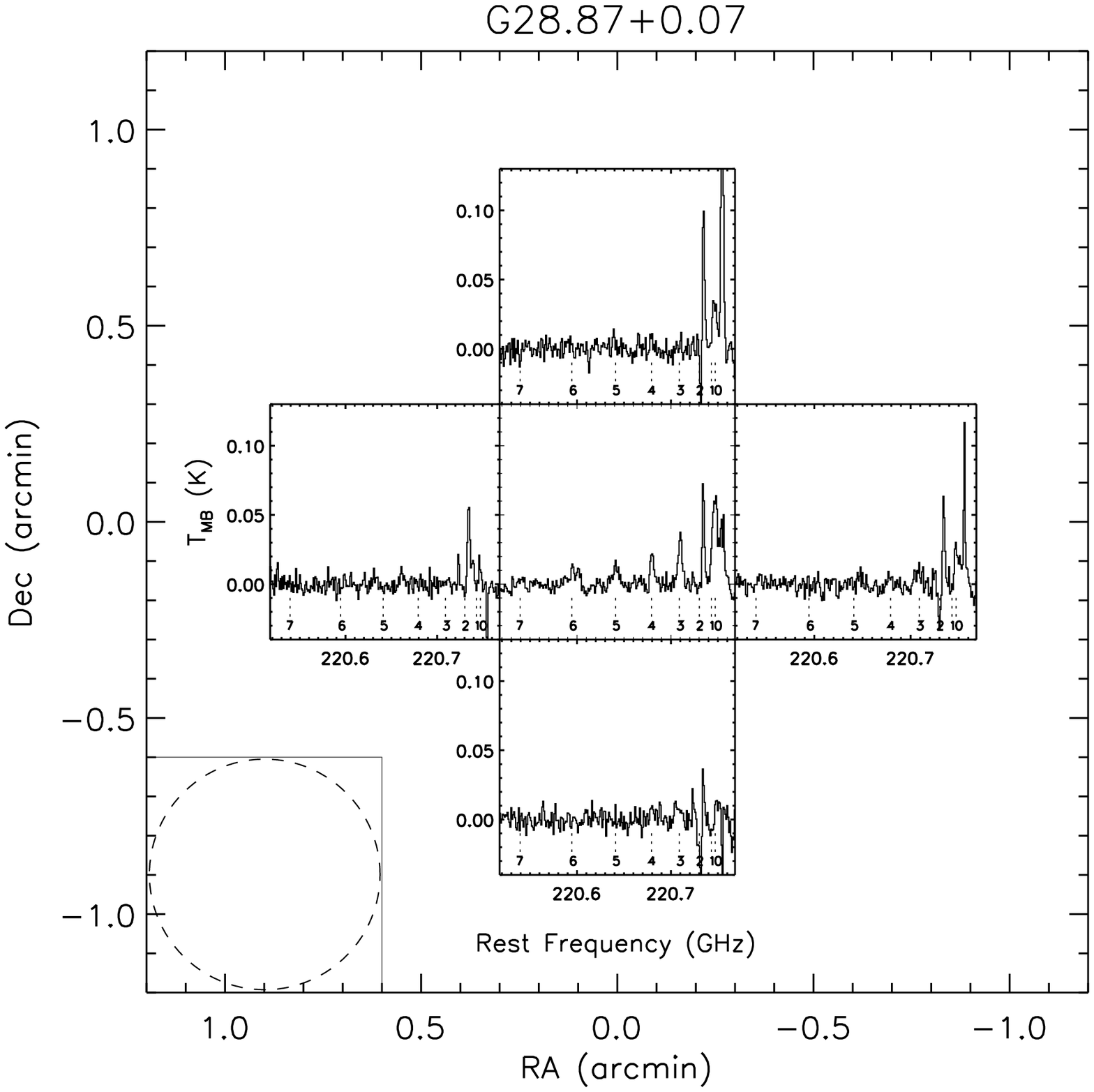,width=0.41\linewidth} \\
\epsfig{file=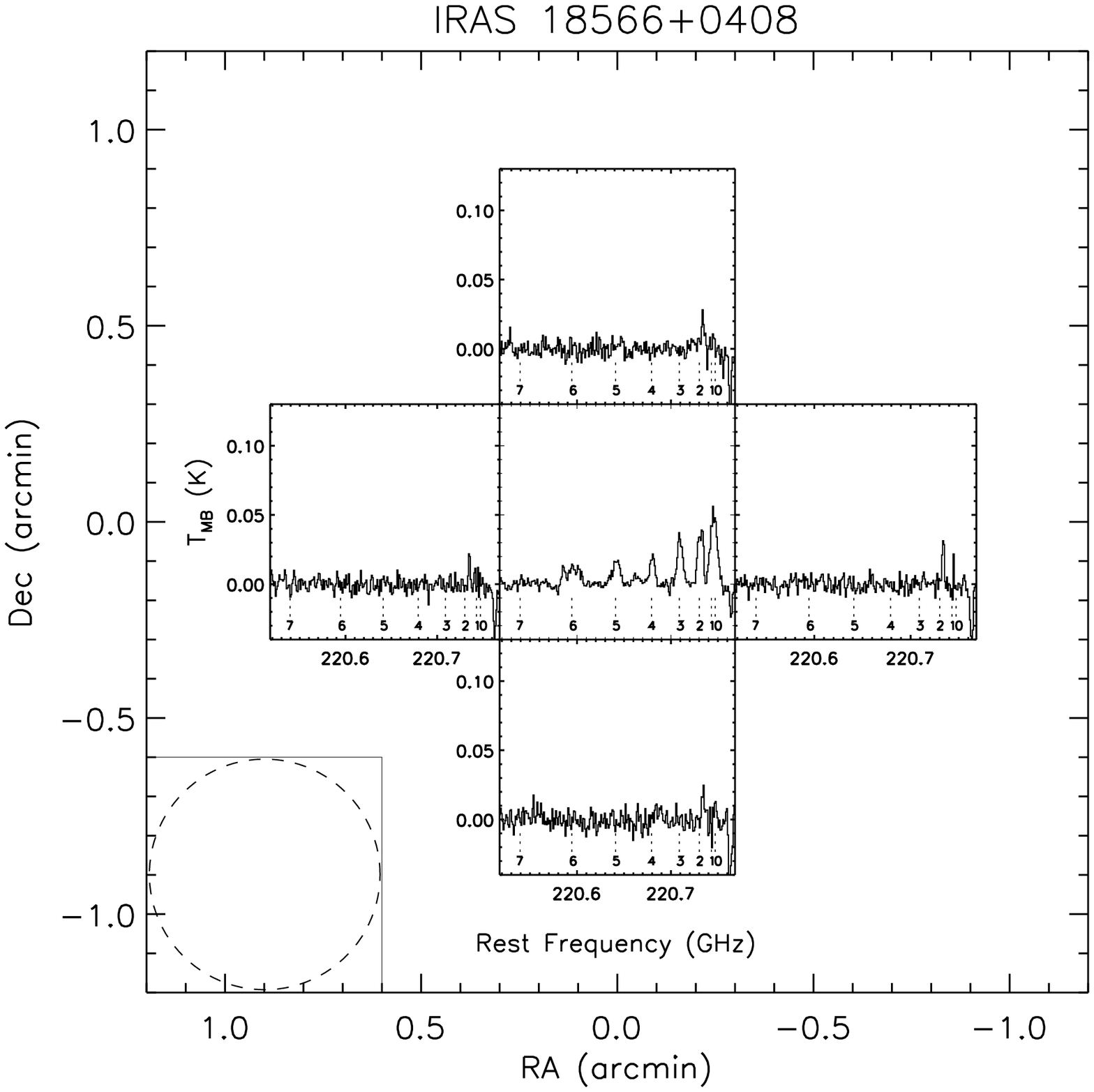,width=0.41\linewidth} &
\epsfig{file=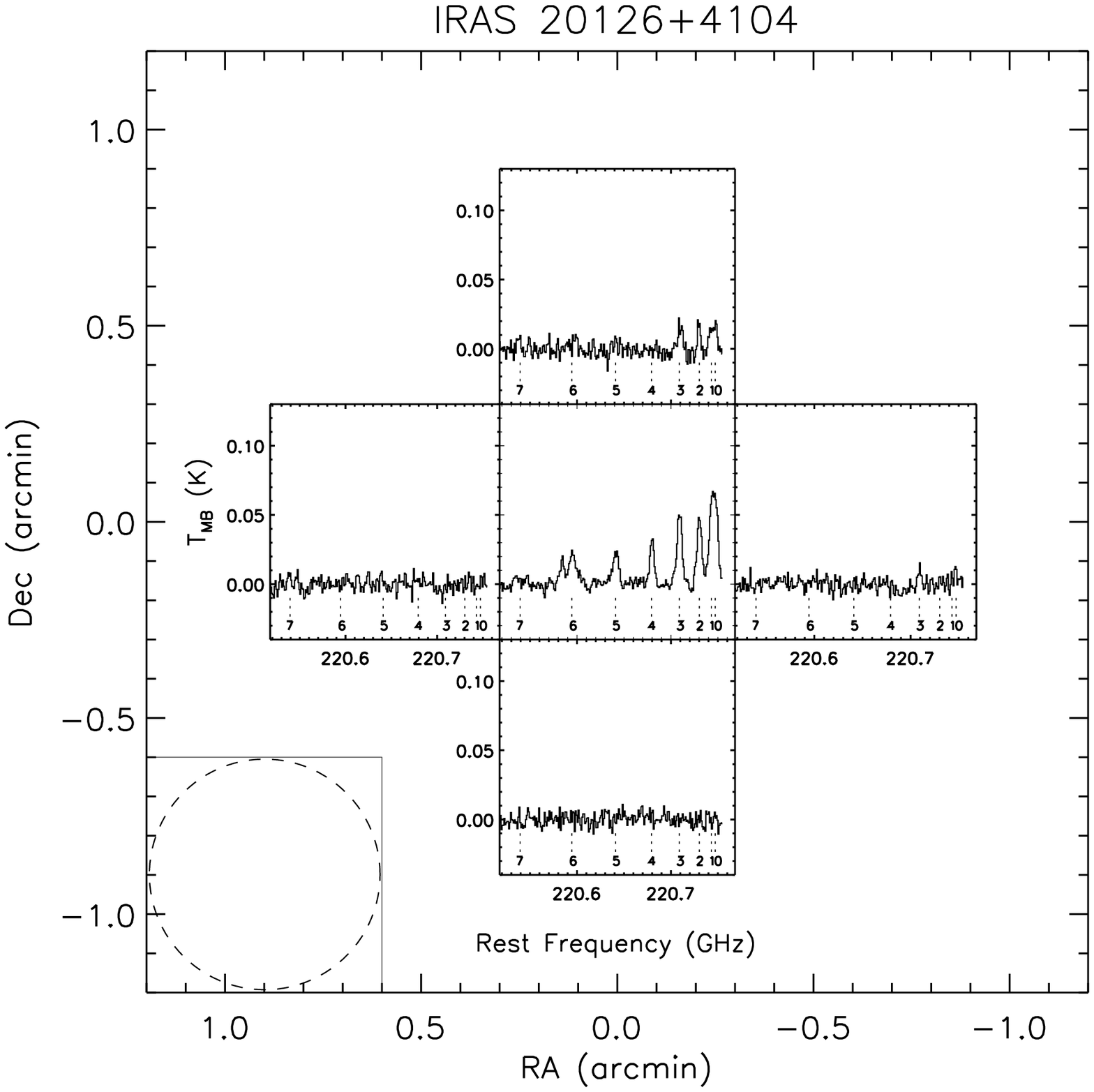,width=0.41\linewidth}\\
\epsfig{file=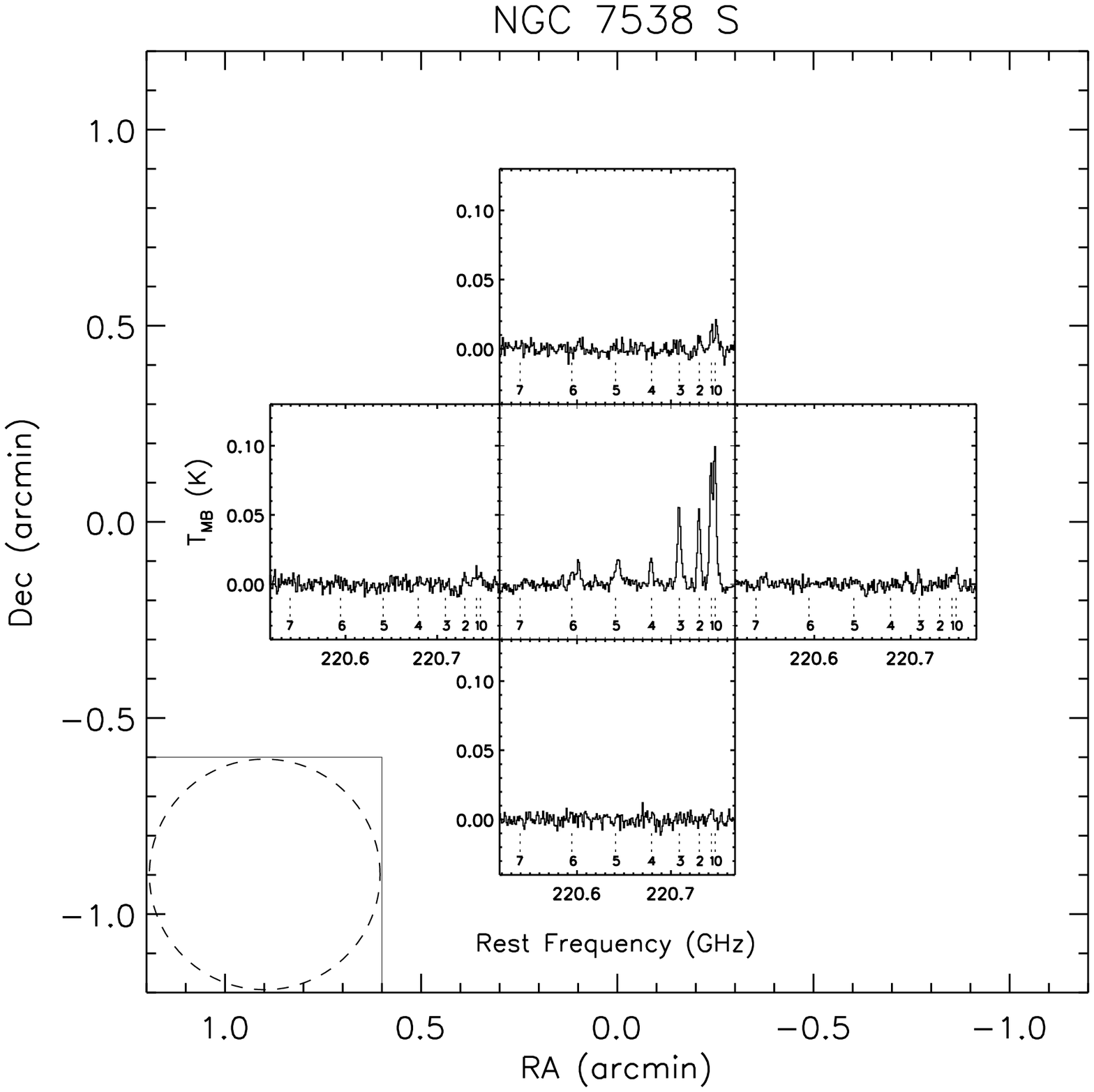,width=0.41\linewidth}
\end{tabular}
\caption{Full beam spaced cross scans towards  5 sources obtained with the SMT$\,$10$\,$m 
telescope (beam size 36$^{\prime\prime}$). The dashed lines below the spectra indicate the position
of the $K-$components of the CH$_3$CN J\,=\,12 $\rightarrow$ 11 transition. The telescope 
beam at FWHM is shown at the bottom left corner.}\label{f3}
\end{figure*}

 In Figure \ref{f3} we show the full beam spaced cross scans that we obtained toward five sources.
 Inspection of the figure shows that the methyl cyanide emission peaks strongly at the
 central position.  Occasionally the lower $K-$components ($K=0-2$) were detected away from the center position
 (IRAS20126+4104, NGC 7538$\:$S), indicating that for these sources the warm gas giving rise to the
 emission is extended on arc-minute scales, but the higher $K-$components ($K> 3$) are always
 found to be compact with respect to our beam.

We used the population diagram technique to derive the rotation temperature and column density for our targets.
This analysis assumes optically thin lines and level populations  described by a Boltzmann distribution
(e.g., \citet{1999ApJ...517..209G}) and furthermore that all lines trace the same volume of gas.
The data are characterized by a linear fit where the 
negative reciprocal of the slope is T$_{rot}$ and  the \emph{y}-intercept is used to infer the  column density (N$_{CH_{3}CN}$)
(see \citet{2005ApJS..157..279A} for a detailed description).   
When calculating the CH$_{3}$CN column density we adopted a gaussian source size of FWHM of $10^{\prime\prime}$.\\

\begin{figure*}
\centering
\includegraphics[scale=0.85]{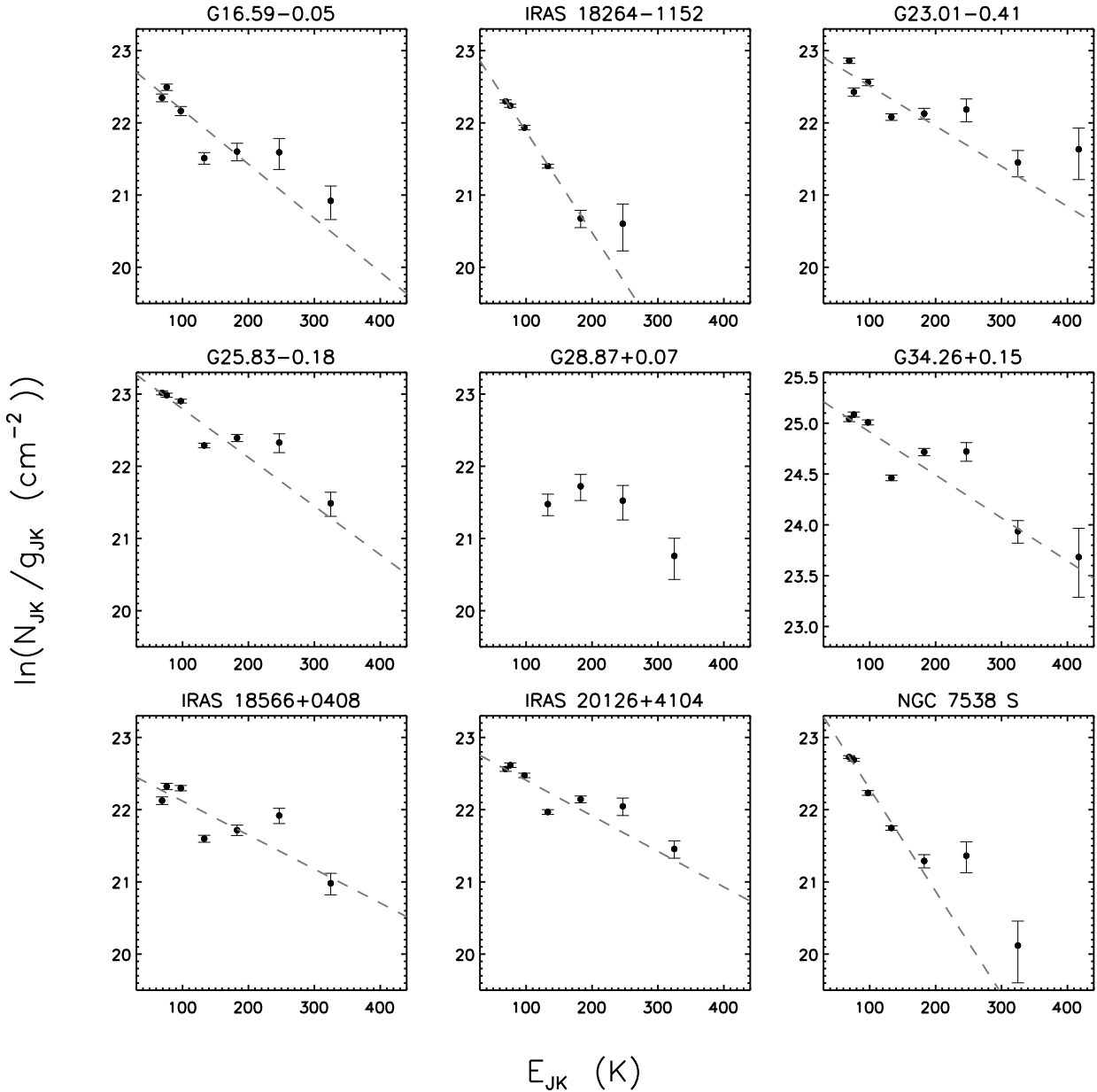}
\caption{The above figure shows Boltzmann plots for the detected sources. No reliable fit could be obtained for
G28.87+0.07. A FWHM of $10^{\prime\prime}$ gaussian source size have been adopted. The error bars are 1$\sigma$.}\label{f4}
\end{figure*}

Figure \ref{f4} shows Boltzmann plots for all detected sources. N$_{JK}$, g$_{JK}$ and  E$_{JK}$
are the column density,  statistical weight  and  upper state energy for the (J,K) state, respectively.
The linear fit was made using only the CH$_3$CN main isotopologue data. 
The results of our population diagram analysis are presented in Table 3. 

\begin{deluxetable}{lcc}		  		  
\tabletypesize{\scriptsize}	
\tablenum{3}				  
\tablecaption{CH$_{3}$CN J\,=12 $\rightarrow$11 Population Diagram Results}	  
\tablehead{					  
\colhead{Source} &				  
\colhead{N$_{CH_{3}CN}$} &				  
\colhead{T$_{rot}$} 
\\	[3pt] 				  
\colhead{} &					  
\colhead{(10$^{14}$\, cm$^{-2}$)} &					  
\colhead{(K)} 	}						  						  
\startdata                                       
G16.59$-0.05$ 			& 0.6(0.2)  	 & 134(12) \\	[3pt] 
IRAS 18264$-1152$        		& 0.30(0.04)     	&   71(2) \\	[3pt] 
G23.01$-0.41$ 			& 1.1(0.3)    	 & 180(14) \\	[3pt] 
G25.83$-0.18$ 			& 1.2(0.2)           & 148(8) \\	[3pt] 
G28.87$+$0.07\tablenotemark{a}  			& \nodata       &   \nodata \\	[3pt] 
G34.26$+$0.15 			& 18.0(3)            & 236(14) \\	[3pt] 
IRAS 18566$+$0408 		& 0.9(0.2)     	 & 212(19) \\	[3pt] 
IRAS 20126$+$4104 		& 1.2(0.2)     	  & 203(13) \\	[3pt] 
NGC 7538$\:$S 			& 0.46(0.06)    	  &    70(2) \\	[3pt] 
\enddata

\tablecomments{A Gaussian source with FWHM of $10^{\prime\prime}$   has been adopted.\label{t3} }
\tablenotetext{a}{No reliable fit was possible.}
\end{deluxetable}

With the low signal-to-noise (S/N) ratio of our spectra,   
detection of the CH$_3$$^{13}$CN isotopologue would indicate optically thick lines.
However, only for G34.26+0.15 we detected  the first three $K-$components of CH$_3$$^{13}$CN  above a $3\sigma$ level. 
The optical depth of the main line $K=2$ (which is less blended)
estimated from the ratio of the line intensities is $\sim 8$,
assuming $^{12}$C/$^{13}$C$=50$. The ratio was calculated based on the abundance variation 
with galacto-centric distance \citep{1994ARA&A..32..191W}. 

For the other targets, no reliable detection of isotopologue lines was made, therefore a detailed analysis correcting for optical
depth effects was not feasible. If the assumption of optically thin conditions is incorrect, the 
derived temperature values are overestimated, and the column density underestimated.
However, optical depths effects will not affect our principal result that relatively hot molecular gas
is present in our targets.

\subsection{Kinematics}

The above analysis of the physical parameters assumed uniform conditions in the emitting gas. This is clearly a simplification, as in
most cases one would expect gradients in the density and temperature. Allowing for non-uniformity in these parameters is not possible with
the present low angular resolution data, however we consider here whether kinematic features can be detected in the CH$_3$CN lines.

Our data  in principle allow us to check for
gradients in line velocity and/or line width. For most of our sources the assumption of central heating, and hence increasing
temperature toward the center of the core is reasonable. Thus, any change of line kinematic parameters for the different $K-$components
which have rapidly increasing excitation energies, would imply radial gradients toward the core center.

To search for radial gradients we have thus obtained independent fits for each line, where
each $K-$component was fit with a Gaussian function independent of the other components. Therefore, 
the FWHM, line position, and intensity of the lines are free parameters in the fit. 
For the main isotopologue the $K=5$ line was fit with three Gaussians when blending with CH$_3$$^{13}$CN $K=0$ and $K=1$ was 
suspected. The $K=6$ line was fit taking into account that its high-frequency wing is blended with the $K=3$ component of CH$_3$$^{13}$CN and
a vibrational line of CH$_{3}$CH$_{2}$OH. The results of these alternative fits are shown in Table 4.

\begin{figure*}
\centering
\includegraphics[scale=0.85]{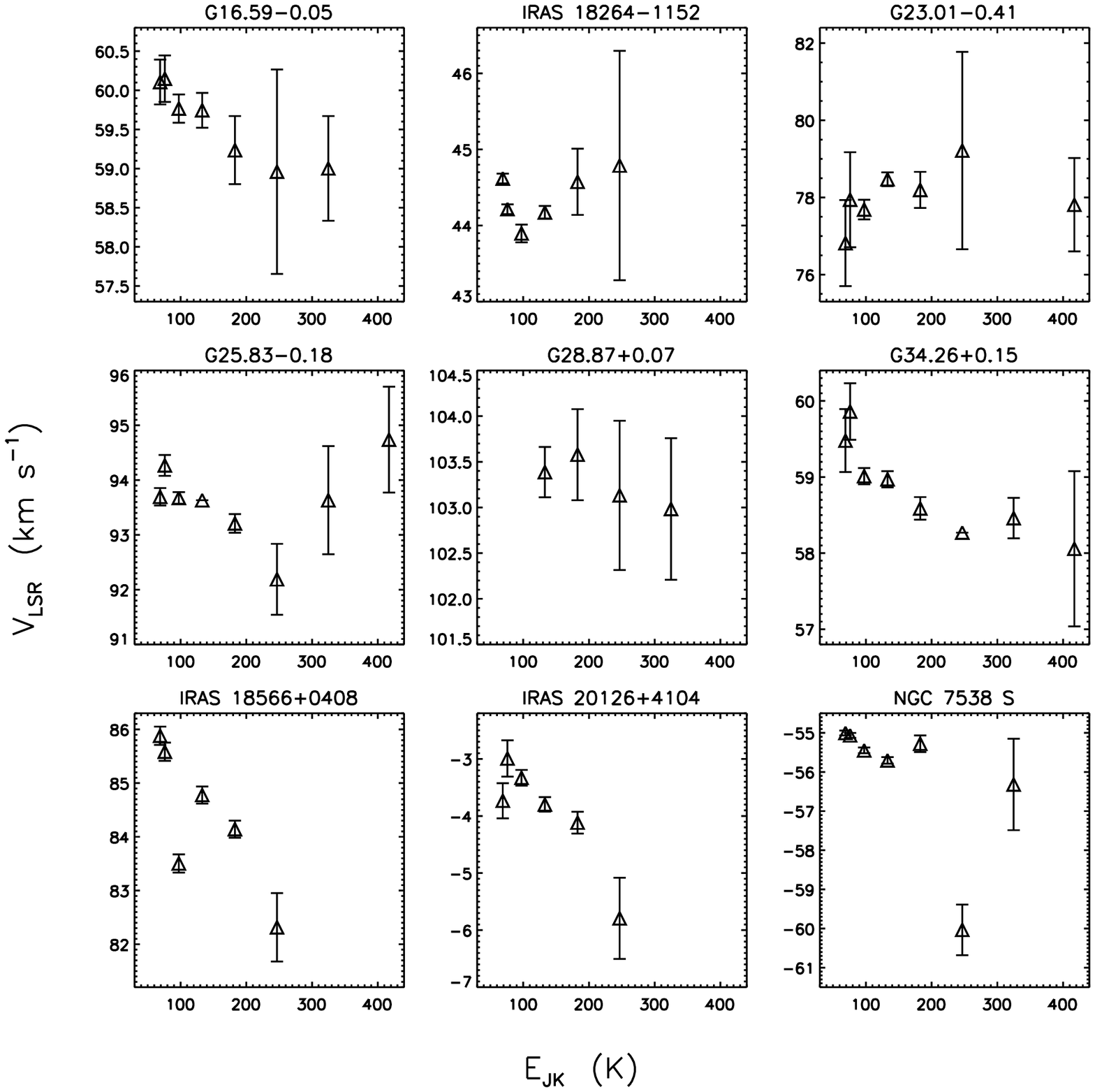}
\caption{Plots of central line velocity vs. upper energy level for all $K-$components which could be fitted
independently. Trends of increasing blueshift with increasing energy are evident for four sources. 
The error bars are 1$\sigma$.}\label{f5}
\end{figure*}

\begin{figure*}
\centering
\includegraphics[scale=0.85]{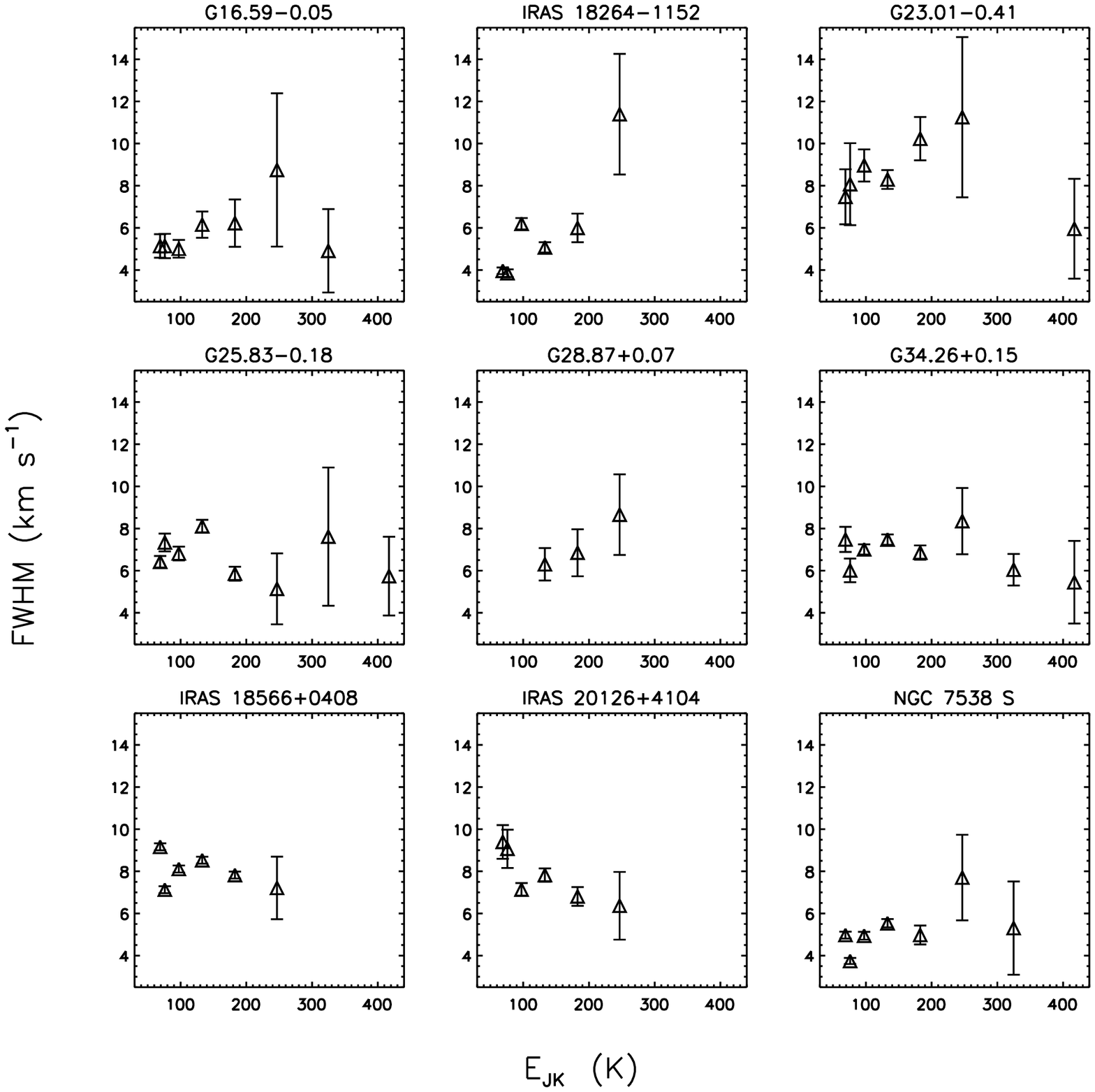}
\caption{Plots of line width vs. upper energy level for all $K-$components which could be fitted
independently. The error bars are 1$\sigma$.}\label{f6}
\end{figure*}

Figures \ref{f5} and \ref{f6} show plots of velocity and line-width as a function of the upper level energy.
From Figure \ref{f5} there appears to be a trend in velocity for G16.59$-$0.05,
G34.26+0.15, IRAS 18566+0408, and IRAS 20126+4104. In all cases the putative velocity gradient is towards lower velocity with
increasing upper state energy. The situation for the line-widths (Figure \ref{f6}) is less clear. Weak evidence for increasing line width towards higher energies is present in
IRAS 18264$-$1152, and G23.01$-$0.41, whereas the opposite trend is seen in IRAS 20126$+$4104. 
We comment further on these findings in the next section.

\section {Discussion}\label{disc_sec}

We have used the 10 m SMT to search for CH$_3$CN emission toward 21 regions of massive star formation.
All of the regions (except for  G34.26$+$0.15) are clearly in an evolutionary phase prior to that of UC or HC HII regions.
We detected nine sources in the CH$_3$CN J\,=\,12 $\rightarrow$ 11 transition. For six of these sources this is the first detection in this transition.
The low detection rate is likely due to the limited sensitivity of
a single dish telescope with the relatively small aperture of $10\,$m.
 
Our cross scans for 5 sources show that the $K\le 3$ are occasionally detected outside the central beam, but the
higher $K-$components are always compact with respect to our angular resolution. This is consistent with a warm (T$\,\leq 50\,$K),
extended envelope which surrounds a much smaller and hotter (T$\, \geq 50\,$K) region near the massive star. This finding is consistent with 
observations of other molecular transitions of lower excitation and critical density (e.g., NH$_3$(1,1)) which probe the more extended envelope.  

Our data thus indicate the presence of gradients in temperature and density. Hence, to obtain precise physical parameters interferometric
observations are necessary. The single dish beam will receive emission from regions of different physical characteristics, and performing the
simple technique of rotation diagram technique corresponds to obtaining average physical properties of the emitting gas. 
Taking also into account a number of additional uncertainties intrinsic to single dish data analyzed with the rotation diagram method (e.g., \citet{2001ApJ...558..194P})
it is difficult to compare our results with the of \citet{1993A&A...276..489O} and \citet{2005ApJS..157..279A}.
Nonetheless we note that our results for the column densities ($\approx 10^{14}\,$ cm$^{-2}$) are quite similar to what was obtained in these studies.
However, our derived temperatures appear to be higher compared to the values of \citet{1993A&A...276..489O}  and \citet{2005ApJS..157..279A}. This might be related to the
different evolutionary state of the target samples. \citet{1993A&A...276..489O} observed many known UCHII regions and the \citet{2005ApJS..157..279A} sample was 
selected from \emph{IRAS} colors typical of UCHII regions, whereas our sample  should represent a pre-UCHII region evolutionary phase. Higher S/N ratio data,
which can put more stringent constraints on the optical depth in the lines will be required to verify this speculation.

We conclude that all detected regions contain molecular gas with 
T$\, > 50\,$K and methyl cyanide column densities of approximately $10^{14}\,$cm$^{-2}$. Assuming a strongly enhanced CH$_3$CN abundance of
$10^{-8}$ the hydrogen column density in these regions on a scale of $10^{\prime\prime}$, is then approximately $10^{22}\,$cm$^{-2}$. These numbers
are consistent with the HMC nature of the detected sources. 

If the core is heated from the center, then 
lower $K-$component emission arises predominantly from outer, and hence cooler
regions and higher $K-$components will dominate in the inner, hotter regions.  
Consequently, velocity gradients in the $K-$ladder components can trace radial dynamics and possible 
kinematic scenarios may be either collapse, outflows, expansion or rotation (e.g, \citet{1997A&A...325..725C}, \citet{2004ApJ...601L.187B}).
We have searched our high resolution spectra for evidence of systematic variation in velocity or 
line-width as a function of excitation energies. 

A small number  of kinematic analyses in CH$_3$CN has been done but often no conclusive
results have been found due to spectral resolution or S/N limitations. For instance,
\citet{1993A&A...276..489O} did not find any reliable trend in their results. On the other hand,
\citet{1999A&A...345..949C} observed IRAS$\,$20126$+$4104 using
the IRAM Plateau de Bure interferometer. These authors reported an increase in the velocity and line-width
toward the center of the core. They suggested that this behavior may be due to rotation of a Keplerian disk
surrounding an embedded massive protostar. Our results for this same region show a decrease in the velocity and the line-width toward the center,
contrary to the results obtained by \citet{1999A&A...345..949C}.
A possible explanation for this discrepancy is the much larger beam of the SMT which will include more extended gas than the
observations of \citet{1999A&A...345..949C}.

We detected an increasing blue-shift, i.e., toward lower velocities with higher $K-$component for four sources.
This trend could be explained by expansion or outflow motion as seen in an optically thick line. 
Another explanation for this behavior could be self absorption of an infalling molecular core
which causes a blue asymmetry in an optically thick line. However with the given S/N ratio of our
spectra we cannot clearly distinguish these scenarios. A further complication that we can not address is the multiple core 
scenario. In this case, the cores have some dispersion of radial velocities that may be unresolved in the beam.
Therefore, to correctly interpret the detected velocity trends, further observations with higher S/N and angular resolution are needed.
\begin{figure*}
\centering
\begin{tabular}{cc}
\includegraphics[scale=0.28]{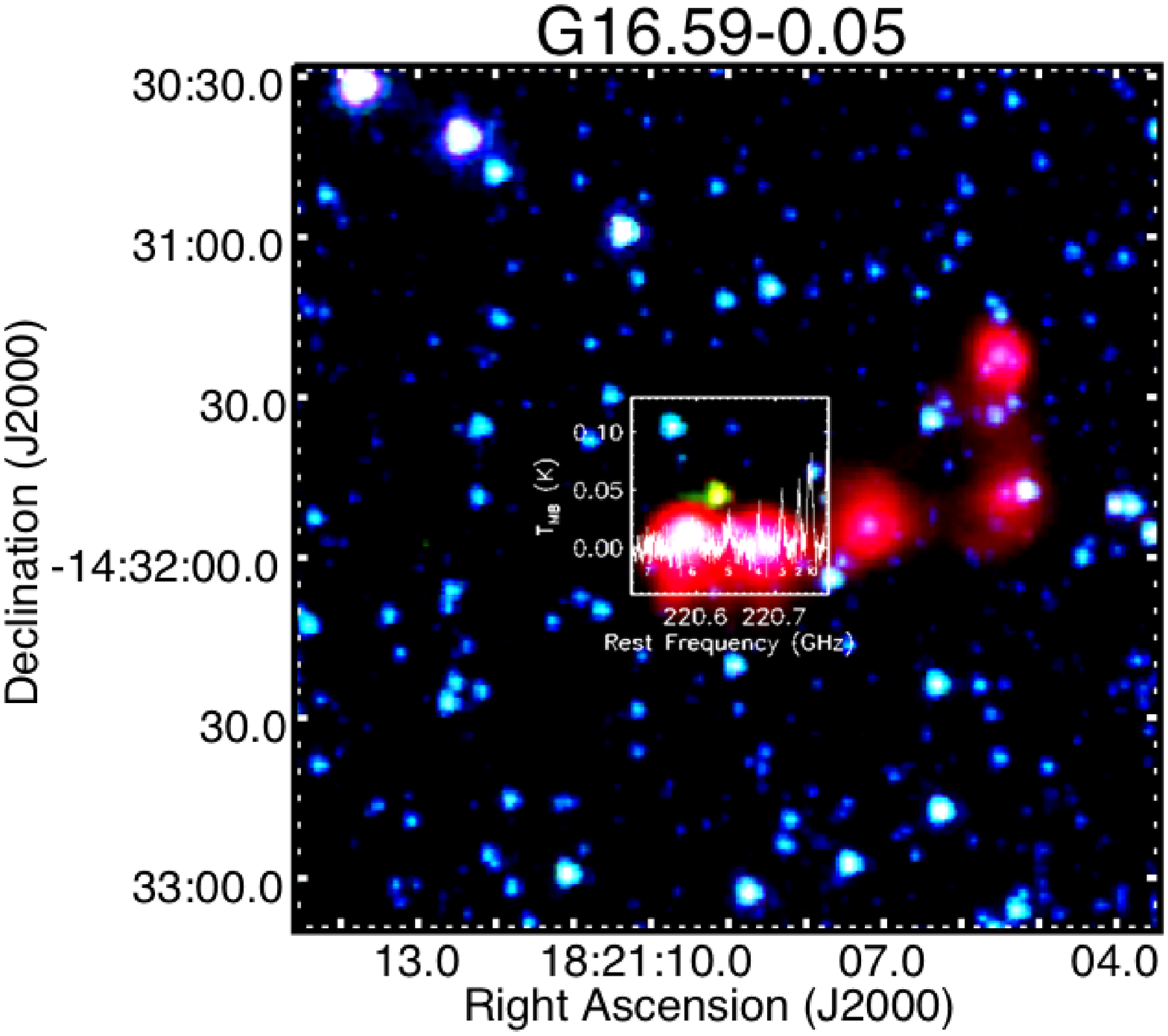} & 
\includegraphics[scale=0.28]{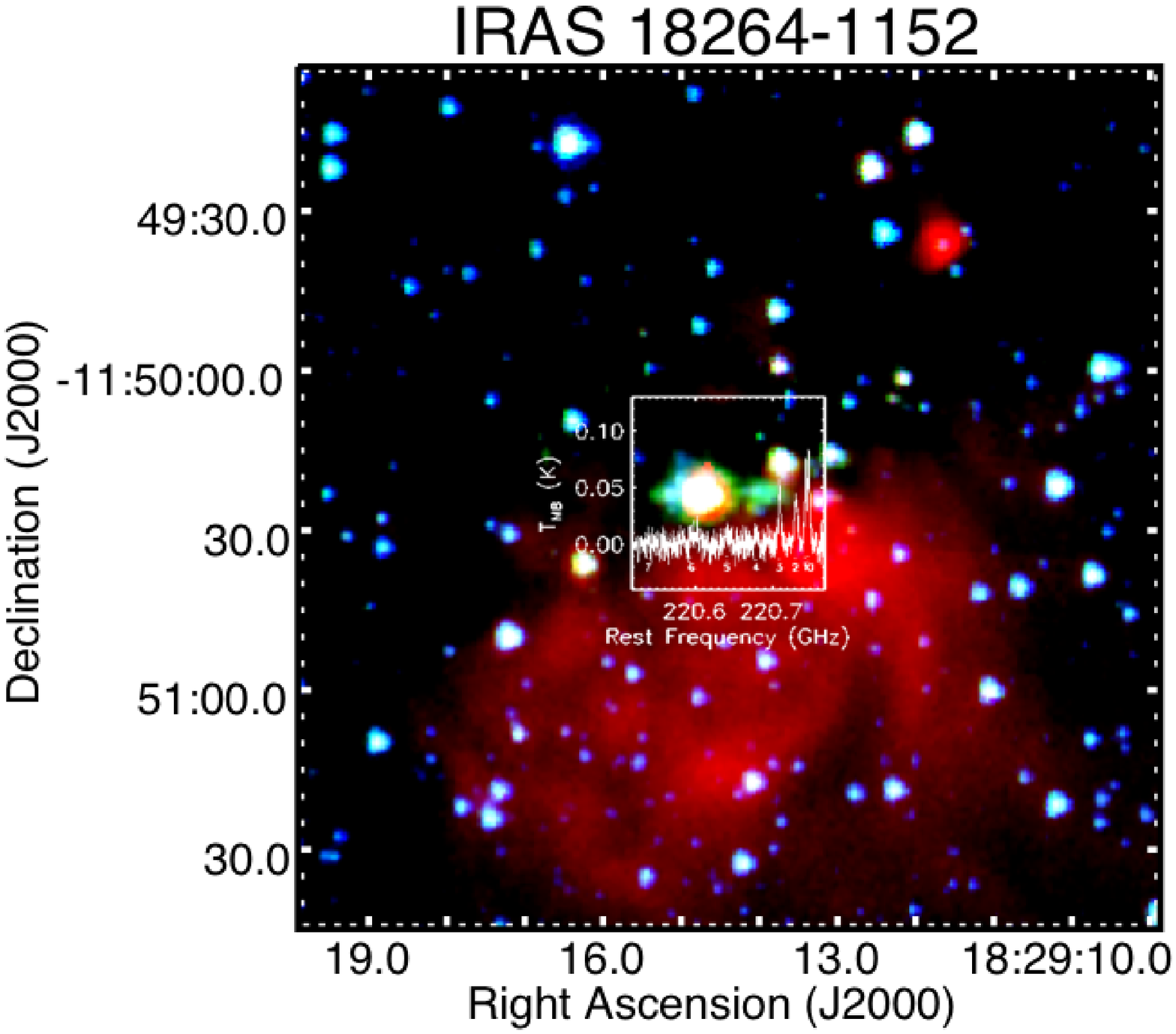}\\
\includegraphics[scale=0.28]{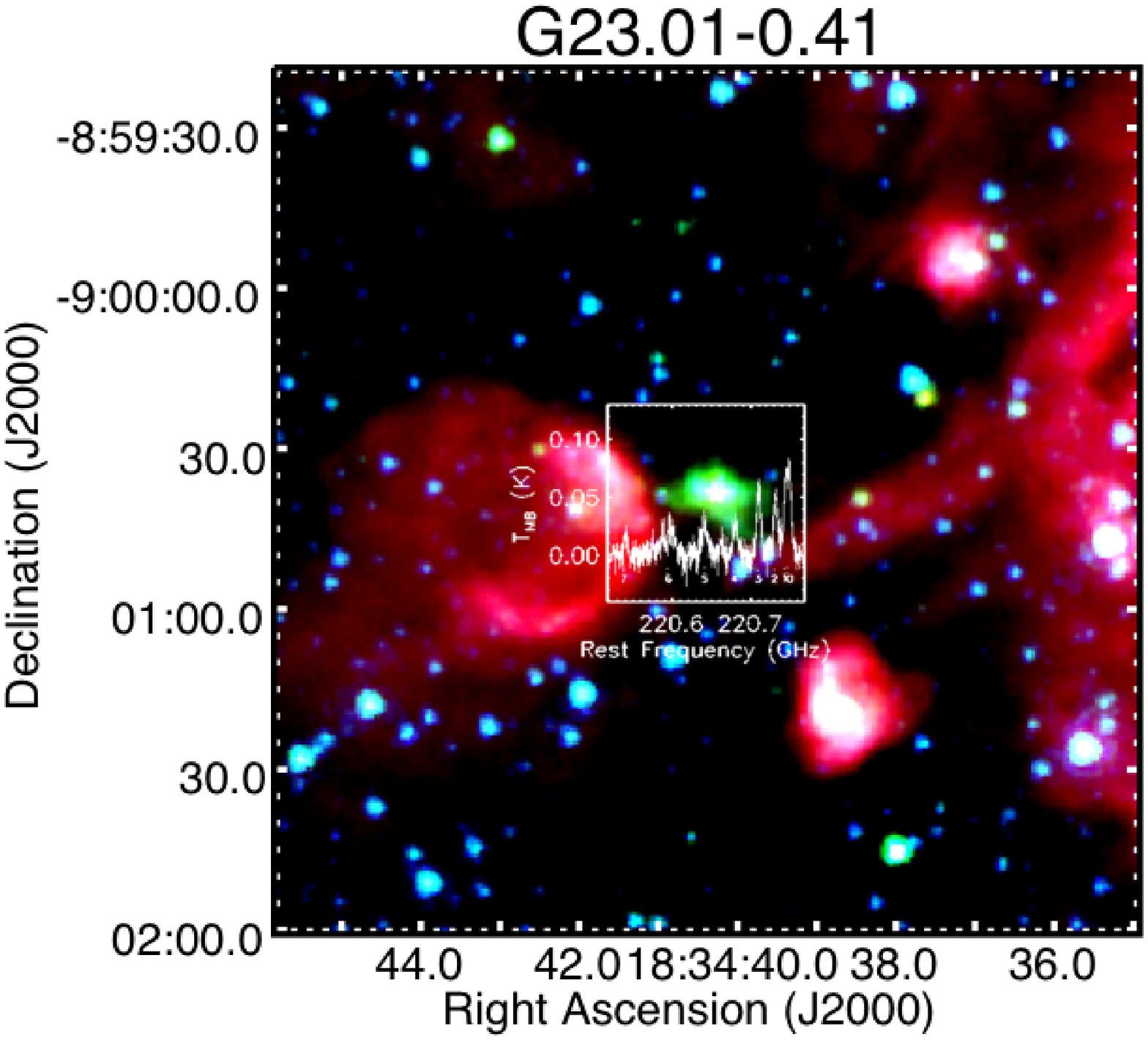} &
\includegraphics[scale=0.28]{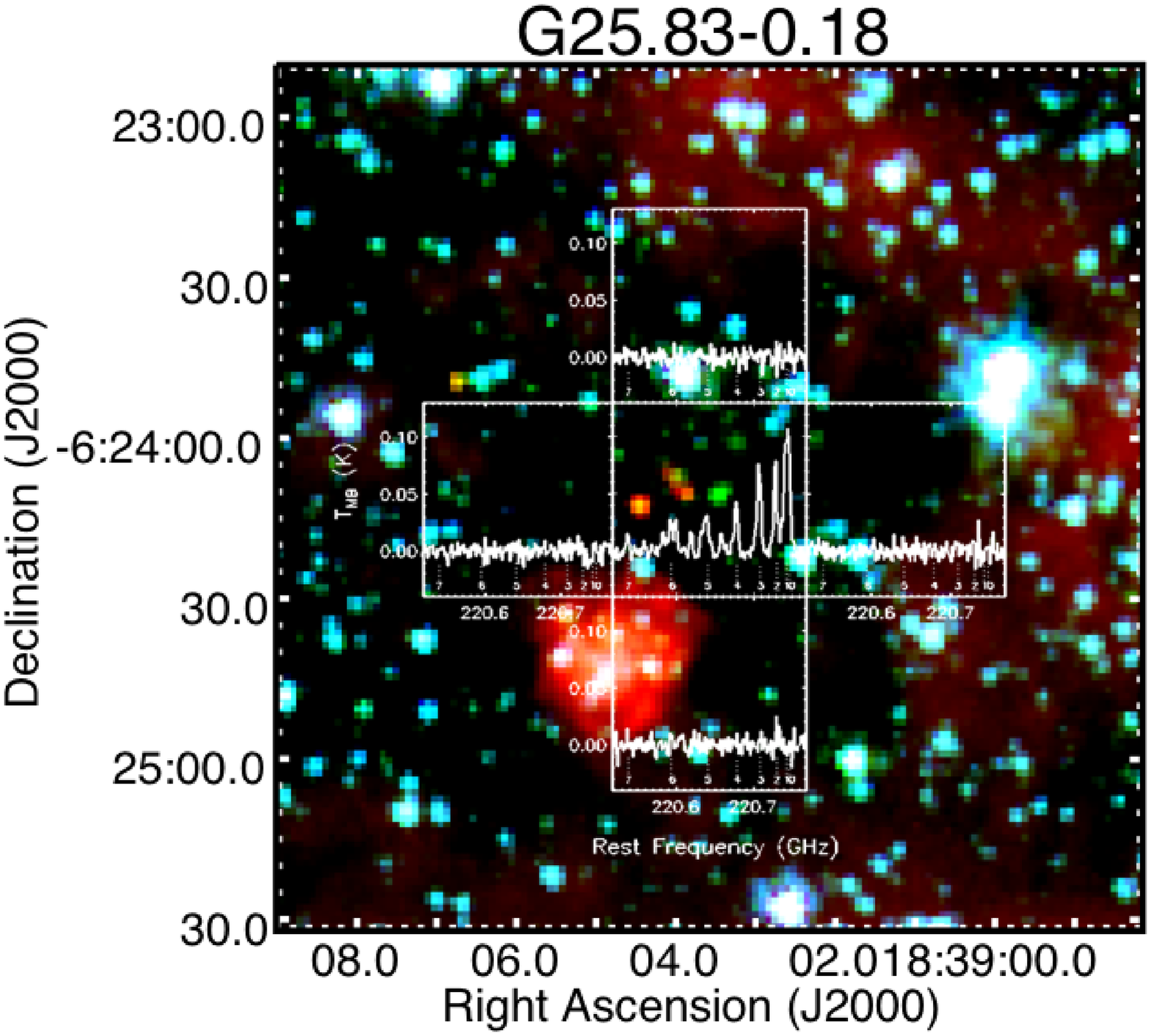} \\
\includegraphics[scale=0.28]{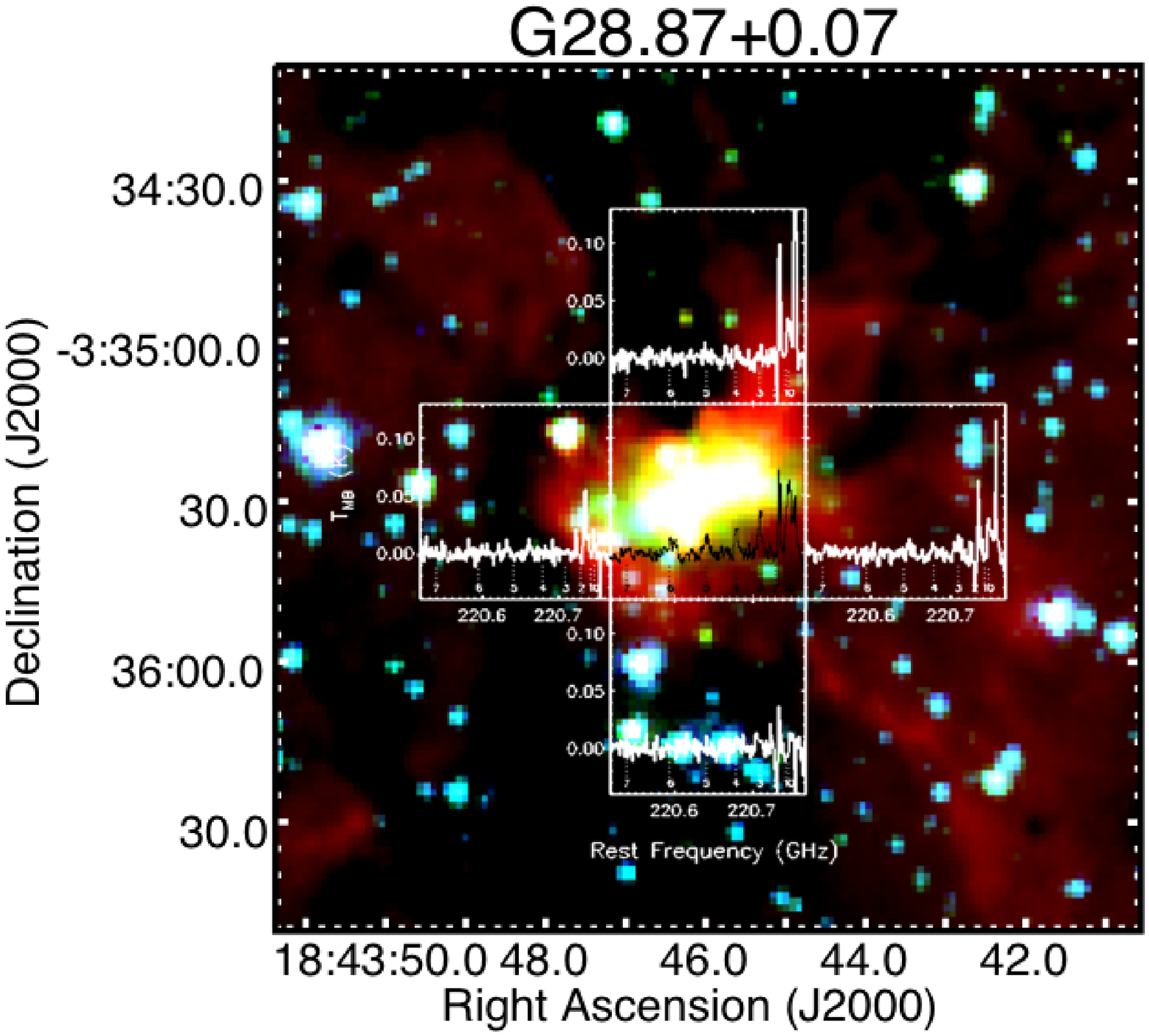} &
\includegraphics[scale=0.28]{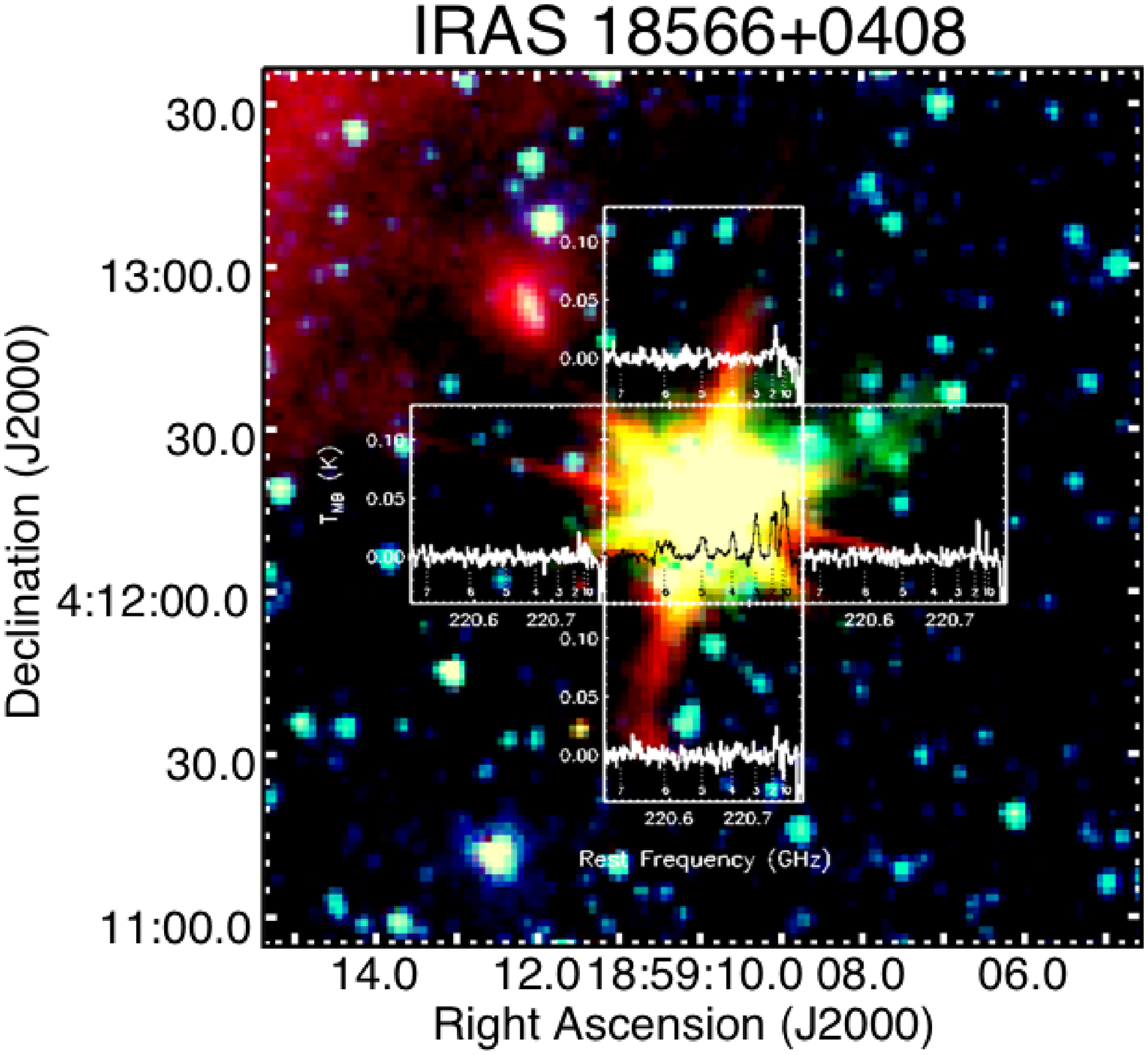}
\end{tabular}

\caption{\emph{SPITZER GLIMPSE}/IRAC three-color (3.6\,$\mu m$-blue, 4.5\,$\mu m$-green and 8.0\,$\mu m$-red) images toward six of our target sources.
The CH$_3$CN J\,=\,12 $\rightarrow$ 11 spectra toward these sources are overlaid on the images. The center of each spectrum is placed at the
corresponding pointing position, 
and the size of each spectrum corresponds to the beam size. }\label{f7}
\end{figure*}

Most of the HMCs observed in this paper are  associated with bright \emph{IRAS} sources, indicating that a
massive object has already heated the surrounding matter.
In Figure \ref{f7} we show mid-infrared images of six sources for which \emph{SPITZER GLIMPSE}/IRAC data are available.
For five out of these six sources there is evidence for extended $4.5\,\mu$m  excess emission (e.g., \citet{2008AJ....136.2391C}).
The presence of an extended $4.5\,\mu$m excess is generally thought to be caused by emission of high
excitation lines from the H$_2$ molecule, and is hence an indicator of shocked gas.
Also, five out of the six sources shown in Figure \ref{f7} have strong mid-IR sources, which clearly shows
that at least these HMCs have already formed massive stars which cause substantial heating of the molecular cores.
One source, G25.83$-$0.18, shows only a dark cloud at the suspected position of the massive protostar. 
This source also harbors one of the few 6 cm H$_{2}$CO masers known \citep{2008ApJS..178..330A}. 
We note that this source shows quite strong CH$_3$CN emission, and since this molecule is thought be effectively
formed only with the presence of an energy source, it is likely that a massive protostar is present in this
dark cloud also. It is tempting then to speculate that due to the different mid-IR properties  G25.83$-$0.18
is in an earlier evolutionary state compared to the other sources which have very bright mid-IR emission. 
However, using the extinction curve of \citet{1990ARA&A..28...37M} we find that hydrogen column densities of $\ge 2\times 10^{23}\,$cm$^{-2}$
could render a possible mid-IR source undetectable. Since hydrogen column densities of this order, and larger, are common in massive
star forming cores, it is conceivable that the different mid-IR appearance of G25.83$-$0.18 is caused by extinction.

\section {Summary}

We have observed a sample of 21 massive proto-stellar candidates  in the CH$_3$CN J\,=\,12 $\rightarrow$ 11 transition using
the 10 m SMT. We detected methyl cyanide in nine of the sources of the sample. 

Rotation temperatures and column densities were estimated using the population
diagram technique. Detected sources have  temperatures  $> $50 K and column densities of $\sim 10^{14}$ cm$^{-2}$,
hence they are consistent with HMC nature.  Higher  spectral resolution is needed in order to understand the structure of 
the kinematics in  molecular cores.

\acknowledgements

 This project was partially supported by the New Mexico Space Grant 
 Consortium.
 PH acknowledges partial support from NSF grant AST-0908901.
SK acknowledges partial support from UNAM DGAPA grant IN101310. 
We thank E. Jordan for help with the observations and initial data reduction.
We thank the anonymous referee for suggestions which improved this
manuscript.
 

%
%
%
%


\clearpage
\LongTables
\tabletypesize{\scriptsize}
\begin{deluxetable*}{@{\extracolsep{-2pt}}lrrrrrrrrrr}
\tablecaption{CH$_3$CN J\,=\,12 $\rightarrow$ 11 Line Parameters}
\tablehead{					  					  		
\colhead{Source} &	
\colhead{V$_{LSR}$} &			  		
\colhead{FWHM} &				  					  			
\multicolumn{8}{c}{$\int$T$_{MB}\,dv$ (K\,km\,s$^{-1}$)} \\		[3pt] 
\cline{4-11}
\colhead{} &
\colhead{(km\,s$^{-1}$)} &
\colhead{(km\,s$^{-1}$)} &
\colhead{$K=0$} &	
\colhead{$K=1$} &
\colhead{$K=2$} &
\colhead{$K=3$} &
\colhead{$K=4$} &
\colhead{$K=5$} &
\colhead{$K=6$} &
\colhead{$K=7$} }						  
						  
\startdata                                       		  
G16.59$-0.05$            & 59.8(0.2)      & 6.2(0.6)        & 0.35(0.02)      & 0.41(0.02)       &0.29(0.02)      & 0.29(0.02)      & 0.15(0.02)      & 0.14(0.03)   &  0.13(0.03)     &       \nodata                \\	[3pt] 	
IRAS 18264$-1152$  & 44.17(0.01) & 5.06(0.08)   & 0.336(0.007)  & 0.314(0.007)  &0.227(0.007)  & 0.257(0.007) & 0.059(0.007)  & 0.05(0.02)  &         \nodata                &       \nodata                 \\		[3pt] 
G23.01$-0.41$            & 78.5(0.2)      & 8.3(0.4)       & 0.59(0.02)        & 0.38(0.02)      &0.42(0.02)      & 0.51(0.02)      & 0.25(0.02)      & 0.25(0.04)   &  0.22(0.04)     &  0.11(0.04)\\	[3pt] 
G25.83$-0.18$            & 93.8(0.1)      & 8.1(0.3)       & 0.69(0.02)       & 0.66(0.02)        &0.60(0.02)      & 0.62(0.02)      & 0.33(0.02)      & 0.29(0.04)   &  0.22(0.04)     &  0.12(0.04) \\		[3pt] 
G28.87$+$0.07           &103.4(0.4)     & 6.3(1.2)            &        \nodata                &         \nodata                 &          \nodata             & 0.28(0.04)      & 0.17(0.03)      & 0.13(0.03)   &  0.11(0.03)     &         \nodata                 \\	[3pt] 
G34.26$+$0.15           &59.0(0.1)       & 7.5(0.2)       & 5.2(0.1)            & 5.4(0.1)           &4.9(0.1)           & 5.5(0.2)          & 3.4(0.1)           & 3.1(0.3)        &  2.6(0.3)          & 0.9(0.3)               \\	[3pt] 
IRAS 18566$+$0408 &84.8(0.2)       & 8.4(0.4)       & 0.28(0.02)      & 0.34(0.02)       &0.33(0.01)       & 0.31(0.02)      & 0.17(0.01)      & 0.19(0.02)   &  0.14(0.02)      &       \nodata                    \\ 	[3pt] 
IRAS 20126$+$4104 &$-$3.8(0.1)   & 7.8(0.3)       & 0.44(0.01)       & 0.46(0.01)       &0.39(0.01)       & 0.45(0.02)     & 0.26(0.01)      & 0.22(0.03)   &  0.22(0.03)     &       \nodata                   \\ 	[3pt] 
NGC 7538$\:$S          &$-$55.69(0.08)& 5.5(0.2)    & 0.52(0.01)      & 0.49(0.01)        &0.31(0.01)      & 0.36(0.01)     & 0.11(0.01)       & 0.11(0.02)   &  0.06(0.02)      &     \nodata                    \\	[3pt] 
\enddata
\label{t2}
\end{deluxetable*}



\begin{center}
\LongTables
\begin{deluxetable}{lcrrrr}
\tabletypesize{\scriptsize}	
\tablenum{4}				  				  
\tablecaption{CH$_3$CN J\,=\,12 $\rightarrow$ 11 Line Parameters  (Free Fits)}
\%tablewidth{0pt}				  						  
\tablehead{					  		
\colhead{Source} &	
\colhead{\# K} &			  		
\colhead{T$_{MB}$} &				  		
\colhead{V$_{LSR}$} &				  			
\colhead{FWHM} &
\colhead{$\int$T$_{MB}\,dv$} \\		[3pt] 	  		
\colhead{} &
\colhead{} &					  		
\colhead{(K)} &					  	
\colhead{(km\,s$^{-1}$)} &	
\colhead{(km\,s$^{-1}$)} &	
\colhead{(K\,km\,s$^{-1}$)} \\	}						  
						  
\startdata        
                               		  
G16.59$-0.05$   & 0 & 0.067(0.006)&60.1(0.3) &5.1(0.6) &0.37(0.05)    \\	[3pt]  	     
                               & 1 & 0.068(0.006)&60.2(0.3) &5.1(0.6) &0.37(0.05)    \\	[3pt] 		     
                               & 2 & 0.049(0.006)&59.8(0.2) &5.0(0.4) &0.26(0.02)  \\	[3pt] 		     
                               & 3 & 0.044(0.006)&59.8(0.2) &6.2(0.6) &0.29(0.02)  \\	[3pt] 	     
                               & 4 & 0.023(0.006)&59.2(0.4) &6.2(1.1) &0.15(0.02)  \\	[3pt] 	     
                               & 5 & 0.018(0.006)&58.9(1.3) &8.8(3.6) &0.17(0.06)  \\	[3pt] 	     
                               & 6 & 0.022(0.006)&59.0(0.7) &4.9(1.9) &0.11(0.03)  \\	[3pt] 		     
\\

IRAS 18264$-1152$& 0 & 0.076(0.006)&44.6(0.1) &  3.9(0.2)  &0.32(0.01)  \\	[3pt] 		     
                                     & 1 & 0.068(0.006)&44.2(0.1) &  3.9(0.2)  &0.28(0.01)  \\	[3pt] 		     
                                     & 2 & 0.038(0.006)&43.9(0.1) &  6.2(0.3)  &0.25(0.01)  \\	[3pt] 		     
                                     & 3 & 0.048(0.006)&44.2(0.1) &  5.1(0.2)  &0.26(0.01)  \\	[3pt] 		     
                                     & 4 & 0.010(0.006)&44.6(0.4) &  5.9(0.7)  &0.07(0.01)  \\	[3pt] 		     
                                     & 5\tablenotemark{a} & 0.009(0.006)&44.8(1.5) & 11.4(2.9)  &0.11(0.03) \\[3pt] 	     
\\
G23.01$-0.41$          & 0 & 0.061(0.006) &76.8(1.1)   &7.5(1.3)      &0.49(0.19)    \\	[3pt] 		    
               		         & 1 & 0.061(0.006)  &77.9(1.2)   &8.1(1.9)     &0.53(0.21)    \\	[3pt] 		    
              		         & 2 & 0.047(0.006)  &77.7(0.3)   &8.9(0.8)     &0.45(0.03)  \\	[3pt] 		     
             		         & 3 & 0.057(0.006)  &78.5(0.2)  &8.3(0.5)      &0.51(0.02)  \\	[3pt] 		     
               		         & 4 & 0.026(0.006)  &78.2(0.5)   &10.2(1.0)   &0.28(0.03)  \\	[3pt] 		     
           		         & 5 & 0.023(0.006)  &79.2(2.6)   &11.3(3.8)   &0.28(0.13)  \\	[3pt] 		     
             		         & 6 & 0.021(0.006)  &77.3\tablenotemark{b}   &10.2\tablenotemark{c}     &0.23(0.05)  \\	[3pt] 	
                 	                  & 7 & 0.016(0.006)  &77.8(1.2)   &5.9(2.4)    &0.10(0.04)  \\	[3pt] 		
\\	                
G25.83$-0.18$         & 0 & 0.096(0.008)&93.7(0.2)   &6.4(0.3)   &0.66(0.04)  \\	[3pt] 		     
                                    & 1 & 0.087(0.008)&94.3(0.2)   &7.3(0.4)   &0.68(0.04)  \\	[3pt] 		     
                                    & 2 & 0.076(0.008)&93.7(0.1)   &6.8(0.3)   &0.55(0.02)  \\	[3pt] 		     
                                    & 3 & 0.072(0.008)&93.6(0.0)   &8.1(0.3)   &0.62(0.02)  \\	[3pt] 		     
                                    & 4 & 0.046(0.008)&93.2(0.2)   &5.9(0.3)   &0.29(0.02)  \\	[3pt] 		     
                                    & 5 & 0.031(0.008)&92.2(0.7)   &5.1(1.7)   &0.17(0.05)  \\	[3pt] 		     
                                    & 6\tablenotemark{d} & 0.025(0.008)&93.6(0.9)   &7.6(3.3)   &0.21(0.06)  \\	[3pt] 		     
                                    & 7 & 0.019(0.008)&94.7(0.9)   &5.7(1.9)   &0.12(0.04)  \\	[3pt] 		 
\\    
G28.87$+$0.07            & 0\tablenotemark{e} &   \nodata         &         \nodata        &          \nodata      &        \nodata    \\	[3pt] 	     
                                    & 1\tablenotemark{e} &   \nodata           &       \nodata          &        \nodata        &      \nodata      \\	[3pt] 	     
                                    & 2\tablenotemark{e} &   \nodata           &       \nodata          &        \nodata       &      \nodata      \\	[3pt] 	     
                                    & 3 & 0.041(0.007)  &103.4(0.3)   & 6.3(0.8)   &0.28(0.03)    \\	[3pt] 		     
                                    & 4 & 0.024(0.007)  &103.6(0.5)   & 6.9(1.1)   &0.18(0.02)    \\	[3pt] 		     
                                    & 5\tablenotemark{a} & 0.017(0.007)  &103.1(0.8)   & 8.7(1.9)   &0.15(0.03)  \\	[3pt] 		     
                                    & 6\tablenotemark{d} & 0.017(0.007)  &102.9(0.8)   &6.82\tablenotemark{c}   &0.13(0.02)  \\	[3pt] 		     
 \\										     						     
G34.26$+$0.15        & 0 & 0.81(0.02)  &59.5(0.4)   &7.5(0.6)   &6.4(0.8)    \\	[3pt] 		     
                                    & 1 & 0.65(0.02)  &59.9(0.4)   &6.0(0.6)   &4.2(0.8)    \\	[3pt] 		     
                                    & 2 & 0.64(0.02)  &59.0(0.1)   &7.0(0.2)   &4.8(0.1)    \\	[3pt] 		     
                                    & 3 & 0.69(0.02)  &58.9(0.1)   &7.5(0.2)   &5.5(0.2)    \\	[3pt] 		     
                                    & 4 & 0.44(0.02)  &58.6(0.2)   &6.9(0.3)   &3.2(0.1)    \\	[3pt] 		     
                                    & 5 & 0.35(0.02)  &58.3(0.0)   &8.4(1.6)   &3.1(0.5)    \\	[3pt] 		     
                                    & 6 & 0.35(0.02)  &58.5(0.3)   &6.1(1.4)   &2.3(0.3)  \\	[3pt] 		     
                                    & 7 & 0.14(0.02)  &58.1(1.0)   &5.5(2.3)   &0.8(0.2) \\	[3pt] 	
\\	     
IRAS 18566$+$0408  & 0 & 0.041(0.007)  &85.9(0.2)   &9.2(0.2)   &0.399(0.003)    \\	[3pt] 	     
                                    & 1 & 0.029(0.007)  &85.6(0.2)   &7.1(0.2)   &0.217(0.003)    \\	[3pt] 		     
                                    & 2 & 0.038(0.007)  &83.5(0.2)   &8.1(0.2)   &0.332(0.003)  \\	[3pt] 		     
                                    & 3 & 0.034(0.007)  &84.8(0.2)   &8.5(0.2)   &0.313(0.003)  \\	[3pt] 		     
                                    & 4 & 0.019(0.007)  &84.1(0.2)   &7.8(0.2)   &0.162(0.003)  \\	[3pt] 		     
                                    & 5 & 0.020(0.007)  &82.3(0.6)   &7.2(1.5)   &0.16(0.03)  \\	[3pt] 		     
                                    & 6\tablenotemark{d} & 0.011(0.007)  &83.0\tablenotemark{b}    &9.5\tablenotemark{c}    &0.11(0.02)  \\	[3pt] 
\\		     
IRAS 20126$+$4104  & 0 & 0.052(0.006)&$-$3.7(0.3)   &9.4(0.8)   &0.53(0.06)  \\	[3pt] 	     
                                    & 1 & 0.046(0.006)&$-$2.9(0.3)   &9.1(0.9)   &0.44(0.06)  \\	[3pt] 		     
                                    & 2 & 0.049(0.006)&$-$3.3(0.1)   &7.1(0.3)   &0.37(0.02)  \\	[3pt] 		     
                                    & 3 & 0.054(0.006)&$-$3.8(0.1)   &7.8(0.3)   &0.45(0.02)  \\	[3pt] 		     
                                    & 4 & 0.033(0.006)&$-$4.1(0.2)   &6.8(0.4)   &0.24(0.01)  \\	[3pt] 		     
                                    & 5 & 0.024(0.006)&$-$5.8(0.7)   &6.4(1.6)   &0.16(0.04)  \\	[3pt] 		     
                                    & 6\tablenotemark{d}  & 0.025(0.006)&-3.3\tablenotemark{b}   &7.9\tablenotemark{c}    &0.21(0.03)  \\	[3pt] 	
\\ 
NGC 7538$\:$S       & 0 & 0.108(0.006)&$-$55.00(0.06)  &4.9(0.2)   &0.57(0.02)  \\	[3pt] 		     
                                    & 1 & 0.097(0.006)&$-$55.07(0.06)  &3.7(0.1)   &0.39(0.02)  \\	[3pt] 		     
                                    & 2 & 0.055(0.006)&$-$55.45(0.07)  &4.9(0.2)   &0.29(0.01)  \\	[3pt] 		     
                                    & 3 & 0.062(0.006)&$-$55.70(0.09)  &5.5(0.2)   &0.36(0.01)  \\	[3pt] 		     
                                    & 4 & 0.019(0.006)&$-$55.3(0.2)  &4.9(0.5)   &0.11(0.01)  \\	[3pt] 		     
                                    & 5 & 0.026(0.006)&$-$60.0(0.7)  &7.7(2.0)   &0.21(0.04)  \\	[3pt] 		     
                                    & 6\tablenotemark{d}  & 0.011(0.006)&$-$56.3(1.2)  &5.3(2.2)   &0.06(0.03)  \\	[3pt]

\enddata
\label{t4}
\tablenotetext{a}{Components K$=0$ and K$=1$ of the isotopologue CH$_3$$^{13}$CN not fitted.}	
\tablenotetext{b}{V$_{LSR}$ artificially constrained in the fit.}
\tablenotetext{c}{FWHM fixed in the fit.}
\tablenotetext{d}{Molecule t$-$CH$_{3}$CH$_{2}$OH not included in fit.}	
\tablenotetext{e}{USB emission is affecting the lines.}	
\end{deluxetable}
\end{center}







\end{document}